\documentclass[letterpaper,11pt]{article}
\usepackage{jheppub}
\usepackage{epstopdf}
\usepackage{slashed}
\usepackage{graphicx} 
\usepackage{soul}
\usepackage{amsmath}
\usepackage{multirow}
\usepackage{tablefootnote}
\usepackage{comment}
\usepackage{color}
\usepackage{caption,subcaption}
\usepackage[normalem]{ulem}

\usepackage[T1]{fontenc}

\usepackage{graphicx}

\usepackage{amsmath}
\usepackage{amsfonts}
\usepackage{amssymb}
\usepackage{cleveref}
\usepackage{color}

\def\be{\begin{equation}}
\def\ee{\end{equation}}
\def\bea{\begin{eqnarray}}
\def\eea{\end{eqnarray}}

\def\<{\langle}
\def\>{\rangle}

\def\slashchar#1{\setbox0=\hbox{$#1$}           
   \dimen0=\wd0                                 
   \setbox1=\hbox{/} \dimen1=\wd1               
   \ifdim\dimen0>\dimen1                        
      \rlap{\hbox to \dimen0{\hfil/\hfil}}      
      #1                                        
   \else                                        
      \rlap{\hbox to \dimen1{\hfil$#1$\hfil}}   
      /                                         
   \fi}          

\def\bea{\begin{eqnarray}}
\def\eea{\end{eqnarray}}
\def\beq{\begin{equation}}
\def\eeq{\end{equation}}

\def\<{\langle}
\def\>{\rangle}

\def\mm{\mu^+\mu^-}
\def\mbb{m_{b\bar b}^{}}

\def\kv{\kappa_V^{}}
\def\kvv{\kappa_{V2}^{}}
\def\kw{\kappa_W^{}}
\def\kz{\kappa_Z^{}}
\def\k3{\kappa_3^{}}
\def\k4{\kappa_4^{}}

\preprint{PITT-PACC-2008}

\title{Electroweak Couplings of the Higgs Boson\\ 
at a Multi-TeV Muon Collider}

\author[a]{Tao Han,}
\author[b]{Da Liu,}
\author[c,d]{Ian Low}
\author[e]{and Xing Wang}

\affiliation[a]{Department of Physics and Astronomy, University of Pittsburgh, Pittsburgh, PA 15217, USA}
\affiliation[b]{Center for Quantum Mathematics and Physics (QMAP),  University of California, Davis, CA 95616, USA}
\affiliation[c]{Department of Physics and Astronomy, Northwestern University, Evanston, IL 60208, USA}
\affiliation[d]{High Energy Physics Division, Argonne National Laboratory, Lemont, IL 60439, USA}
\affiliation[e]{Department of Physics, University of California-San Diego, 
La Jolla, CA 92093, USA}

\abstract{
We estimate the expected precision at a multi-TeV muon collider for measuring the Higgs boson couplings with electroweak gauge bosons, $HVV$ and $HHVV\ (V=W^\pm,Z)$, as well as the trilinear Higgs self-coupling $HHH$. At very high energies both single and double Higgs productions rely on the vector-boson fusion (VBF) topology.
The outgoing remnant particles have a strong tendency to stay in the very forward region, 
leading to the configuration of the ``inclusive process'' and making
 it difficult to isolate $ZZ$ fusion events from the $WW$ fusion. In the single Higgs channel, we perform a maximum likelihood analysis on $HWW$ and $HZZ$ couplings using two categories: the inclusive Higgs production and the 1-muon exclusive signal.
In the double Higgs channel, we consider the inclusive production and study the interplay of the trilinear $HHH$ and the quartic $VVHH$ couplings, by utilizing kinematic information in the invariant mass spectrum. We find that at a centre-of-mass energy of 10 TeV (30 TeV) with an integrated luminosity of 10 ab$^{-1}$ (90 ab$^{-1}$), one may reach a 95\% confidence level sensitivity of 
0.073\% (0.023\%) for $WWH$ coupling, 
0.61\% (0.21\%) for $ZZH$ coupling,
0.62\% (0.20\%) for $WWHH$ coupling, and
5.6\% (2.0\%) for $HHH$ coupling. 
For dim-6 operators contributing to the processes, these sensitivities could probe the new physics scale $\Lambda$ in the order of $1-10$ ($2-20$) TeV at a 10 TeV (30 TeV) muon collider. 
}

\begin{document}

\maketitle

\noindent
\section{Introduction}

The discovery of the Higgs boson at the CERN Large Hadron Collider (LHC) opens a new avenue in particle physics. On the one hand, the existence of the Higgs boson completes the particle spectrum in the Standard Model (SM) and provides a self-consistent mechanism in quantum field theory for mass generation of elementary particles. On the other hand, the SM does not address the underlying mechanism for the electroweak symmetry breaking (EWSB) and thus fails to understand the stability of the weak scale with respect to the Planck scale. In order to gain further insight for those fundamental questions, it is of high priority to study the Higgs boson properties to high precision in the hope to identify hints for new physics beyond the SM.

In the SM, the Higgs sector is constructed from a complex scalar doublet $\Phi$. After the EWSB, the neutral real component is the Higgs boson excitation $H$ and the other three degrees of freedom become the longitudinal components of the massive gauge bosons. As such, studying the Higgs-gauge boson couplings would be the most direct probe to the underlying mechanism of the electroweak symmetry breaking. 
After the EWSB, the Higgs sector can be parameterized as 
\begin{eqnarray}
    \label{eq:kappa}
    \mathcal{L}& \supset& \left(M_W^2W^+_\mu W^{-\mu}+{1\over 2}M_Z^2Z_\mu Z^\mu\right) \left( \kv  {2H\over v} + \kappa_{V_2} {H^2\over v^2} \right) -{m_H^2\over 2 v} \left( \kappa_{3}^{} H^3 + {1\over 4v} \kappa_{4}^{} H^4 \right),~~~~
\end{eqnarray}
where $v=246$ GeV is the vacuum expectation value of the Higgs field and $\kappa_i =1$ for the SM couplings at tree-level. This ``$\kappa$-scheme'' is a convenient phenomenological parameterization of  deviations from the SM expectations, which is suitable for the exploratory nature of the present study. Here it is made implicit that $\kappa_V=\kappa_W=\kappa_Z$. This is the prediction of the tree-level custodial SU(2) invariance \cite{Sikivie:1980hm}, which is an accidental symmetry of the SM. This has been verified to a good accuracy by precision EW measurements \cite{ALEPH:2005ab}. Nevertheless, in our fit we wish to be more general and will not be assuming a correlated $\kappa_W$ and $\kappa_Z$.

A fully consistent and theoretically-sound framework would utilize effective field theories (EFT), by augmenting the SM Lagrangian with higher dimensional operators from integrating out the heavier states \cite{Grzadkowski:2010es}. 
While a systematic account for the effects of the higher dimensional operators is much more involved and beyond the scope of the current work, we would like to consider the  following two operators for the purpose of illustration \cite{Barger:2003rs,Giudice:2007fh} 
\begin{eqnarray}
\begin{split}
     {\mathcal{O}}_H =\frac{c_H}{2\Lambda^2} 
        \partial_\mu(\Phi^\dagger\Phi) \partial^\mu(\Phi^\dagger\Phi)\ , \quad
         {\mathcal{O}}_6 = -\frac{c_6 \lambda}{\Lambda^2} (\Phi^\dagger\Phi)^3 \ ,
        \end{split}
    \label{dim6text}
\end{eqnarray}
where $\Lambda$ is the cutoff scale where new physics sets in, and 
$\lambda$ is the quartic coupling parameter  in front of $(H^\dagger H)^2$ term in the SM Higgs potential. At the dimension-six level these are the two operators that are most relevant for our study. An additional operator, $\Phi^\dagger \Phi (D_\mu \Phi)^\dagger (D^\mu\Phi)$, can be removed by a suitable field-redefinition \cite{Giudice:2007fh}. The resulting shifts $\Delta\kappa_i \equiv \kappa_i-1$ in Eq.~(\ref{eq:kappa}) are\footnote{Interestingly, in most cases there is a positivity constraint on $c_H>0$, thereby reducing the $VVH$ and $VVHH$ coupling strengths \cite{Low:2009di}.}
\begin{eqnarray}
\begin{split}
\Delta \kv &= -{c_H\over 2} {v^2\over \Lambda^2} \ , &  \Delta \kvv &= - 2c_H {v^2\over \Lambda^2} ,\\   
 \Delta \kappa_3 &\approx -{3c_H\over 2} {v^2\over \Lambda^2} + {c_6} {v^2\over\Lambda^2}\ , \quad &\Delta \kappa_4& \approx  - \frac{25}{9} c_H {v^2\over \Lambda^2} + 6 c_6 {v^2\over \Lambda^2}, 
\end{split}
\label{eq:cs}
\end{eqnarray}
We see that deviations in the $VVH$ and $VVHH\ (V=W^\pm,Z)$ couplings are correlated and controlled by the same operator ${\cal O}_H$. However, the precision we are expecting is high and could potentially be sensitive to effects of dimension-8 operators, in which case the correlation may be modified. 
 On the other hand, the Higgs trilinear self-coupling $\kappa_3$ is among the most important interactions to be tested in the Higgs sector -- it governs the shape of the Higgs potential and, consequently, the nature of the electroweak symmetry breaking. In addition, $\kappa_3$ controls the strength of the  electroweak phase transition, which is important for understanding the cosmological evolution of the early universe as well as the origin of the observed matter-anti-matter asymmetry in the current unverse \cite{Zhang_1993,Grojean_2005,Gan_2017}.  Precise measurements of these couplings will provide insights on how nature works at the shortest distance scale ever probed by mankind. Needless to say, should deviations from the SM predictions be observed, it would completely revolutionize  our understanding of the physical laws of nature.

With the great success of the LHC program, we have achieved the measurement of the   $VVH$ to  ${\cal O}(5\%)$ accuracy \cite{Aad:2019mbh,Sirunyan:2018koj}, which will  be further improved by roughly a factor of two with the high-luminosity LHC upgrade \cite{Cepeda:2019klc}. In $e^+e^-$ collisions at the International Linear Collider (ILC) \cite{Asner:2013psa,Tian:2013yda}, the proposed Higgs factories \cite{Abada:2019zxq,CEPCStudyGroup:2018ghi,An_2019} and the CLIC \cite{Robson:2018zje, Roloff:2019crr}, sub-percent level accuracies for $WWH$ of ${\cal O}(0.6\% - 1.2\%)$ and $ZZH$ 
of ${\cal O}(0.2\% - 0.5\%)$ could be achievable. 
However, the trilinear $HHH$ and quartic $VVHH$ couplings are still difficult to measure to an informative level without a very high energy collider \cite{Barger:1988kb,Kilian:2018bhs}. At a 100 TeV hadron collider such as the SPPC or FCC$_{hh}$, the trilinear Higgs self-coupling could be measured with  ${\cal O}(5\%)$ uncertainty \cite{Benedikt:2018csr,CEPC-SPPCStudyGroup:2015csa}. 
Recently, an attempt was made to determine the quartic Higgs self-coupling at a high-energy muon  collider \cite{Chiesa:2020awd}.
In the EFT language, the precision to which one could measure the Higgs couplings can be translated into constraints on the scale suppressing dimension-6 operators, which is indicative of the scale where new physics becomes important. A figure of merit is when $\Lambda \sim 1$ TeV which, generally speaking, would induce a corresponding deviation in the Higgs couplings of the order \cite{Gori:2013mia}
\be
{\cal O}\left(\frac{v^2}{\Lambda^2}\right) \sim {\cal O}(5\%) \qquad {\rm for} \quad \Lambda \sim 1 \ \ {\rm TeV} \ .
\ee
Therefore, in order to probe new physics scale above 1 TeV, it is important to be able to reach a precision level of 5\% or less. In addition, in a lepton collider a truly model-independent determination of the trilinear $HHH$ coupling requires simultaneously measuring the 4-point $VVHH$ coupling, which is difficult to access at low energies and without sufficiently high statistics.

Recently, there has been a renewed interest to consider a muon collider with a very high centre-of-mass (CM) energy in the tens of TeV \cite{Delahaye:2019omf,Long:2020wfp,Buttazzo:2018qqp,Costantini:2020stv,Han:2020uid,Capdevilla:2020qel}. While the previous discussions for a muon collider were focused on a Higgs factory operating at the SM Higgs resonance \cite{Barger:1995hr,Barger:1996jm}, a collider operating at a multi-TeV regime would certainly lead us to a new territory at the energy frontier.  
 Such a multi-TeV  muon collider  offers a unique opportunity to probe the electroweak couplings of the Higgs boson, including $VVH, HHH$ and $VVHH$ couplings. The possible CM  energy under discussion ranges from 3 TeV to 30 TeV, with a representative benchmark target at 10 TeV or higher. Very high luminosities are also envisioned, with the scaling relation as \cite{Delahaye:2019omf}
\beq
{\rm Lumi.} > { {\rm 5~years} \over {\rm time} }\ \left({ \sqrt s \over 10~{\rm TeV} } \right)^2\ 2 \cdot 10^{35}\ {\rm cm}^{-2}  {\rm s}^{-1} .
\eeq
This will yield to an integrated luminosity of ${\cal O}(10)$ ab$^{-1}$ at $\sqrt s = 10$ TeV and ${\cal O}(90)$ ab$^{-1}$ at $\sqrt{s} = 30$ TeV, 
which would take us to a remarkable new energy frontier, and offer great potential to study the Higgs boson, and the Nature in general at an unprecedented short-distance scales. In this paper, we would like to explore the Higgs physics and examine the accuracies for the electroweak couplings of the Higgs boson  at the future high-energy muon collider.

The rest of the paper is organized as follows. 
We first present the Higgs boson production rates via various production mechanisms at a high-energy muon collider in Sec.~\ref{sec:Higgs}. We then evaluate the statistical accuracy achievable to determine the $HVV$ couplings in Sec.~\ref{sec:VVH}. Foremost, we show the improvement for the precision measurement on the triple Higgs boson coupling as well as the $VVHH$ coupling in Sec.~\ref{sec:HH}. We summarize our results and conclude in Sec.~\ref{sec:Sum}. 

\section{Higgs Boson Production at a High-energy Muon Collider}
\label{sec:Higgs}

\begin{figure}[tb]
\centering
\includegraphics[width=0.25\textwidth]{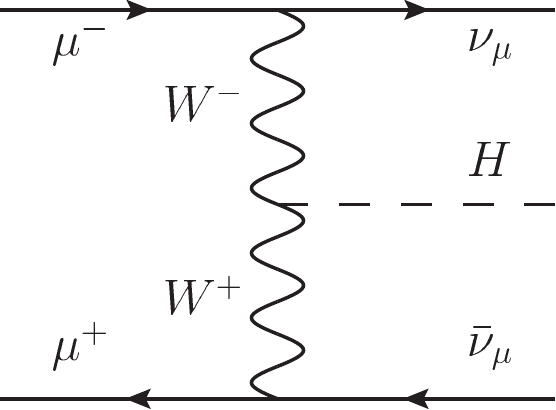}\\
\caption{VBF production of a single Higgs boson at a high energy muon collider via $WW$ fusion. For $ZZ$ fusion, replace the $W$ propagator by the $Z$ propagator and the outgoing neutrinos by muons.}
\label{fig:VBF}
\end{figure}

\begin{figure}[tb]
\centering
\begin{subfigure}{0.25\textwidth}
\includegraphics[width=\textwidth]{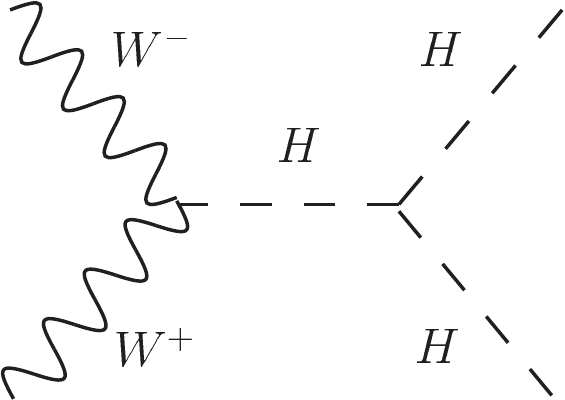}
\caption{}
\label{fig:feyn_s}
\end{subfigure}
\hspace{0.4cm}
\begin{subfigure}{0.25\textwidth}
\includegraphics[width=\textwidth]{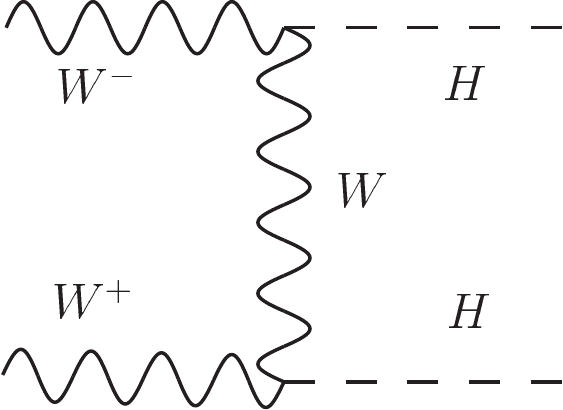}
\caption{}
\label{fig:feyn_t}
\end{subfigure}
\hspace{0.4cm}
\begin{subfigure}{0.25\textwidth}
\includegraphics[width=\textwidth]{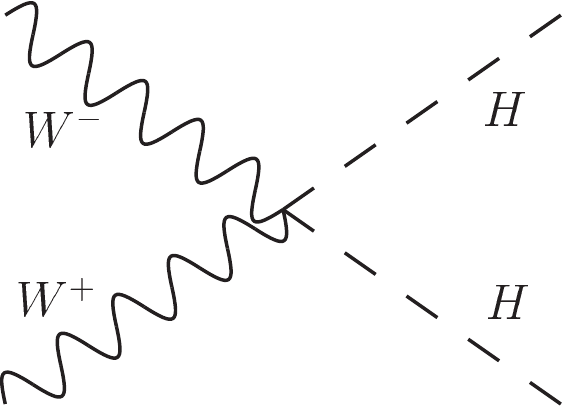}
\caption{}
\label{fig:feyn_4}
\end{subfigure}
\caption{Double Higgs production at a high energy muon collider via $WW$ fusion. The production goes through the VBF topology, as in Fig.~\ref{fig:VBF}.
}
\label{fig:Feyn}
\end{figure}

The Higgs boson couples predominantly to heavier particles. The production of a Higgs boson thus involves  other heavy particles in the SM. At high energies, gauge bosons will copiously radiate off the colliding beams.
Therefore, the vector boson fusion (VBF) mechanism are the dominant source for the Higgs boson  production at a high-energy muon collider \cite{Costantini:2020stv,Han:2020uid}. 
The production processes involving the Higgs boson at a high-energy muon collider include 
\beq
\mm\quad \stackrel{\rm VBF}{\longrightarrow}\quad H,\ ZH,\ HH\quad {\rm and}\quad t\bar t H \ ,
\eeq
which are all dominantly from the VBF processes. 
We list the production cross sections in Table \ref{tab:xsecH} for those Higgs production processes with a few representative benchmark energy choices. Cross sections are computed using the package {\tt MadGraph} \cite{Alwall:2014hca}. Recently it has been advocated that, in high energy collisions, it may be appropriate to adopt the approach of electroweak parton distribution functions (EW PDF)  \cite{Han:2020uid} to resum the potentially large collinear logarithms at high scales. For the processes under consideration, the difference is insignificant since the single Higgs production is set by a low scale $m_H$, while the Higgs pair production $HH$ is dominated by the longitudinal gauge boson fusion ($W_LW_L$), that has no scale dependence at the leading order. 

\begin{table*}[tb]
\centering
\begin{tabular}{|c|c|c|c|c|c|}
\hline
   $\sqrt s$ (TeV)   & {3}    & {6} & {10} & {14} & {30} \\
    benchmark lumi (ab$^{-1}$)   & {1}    & {4} & {10} & {20} & {90} \\
   \hline
  $\sigma$  (fb): $ WW \to H $ & 490 &  700        &  830         & 950           &  1200   \\
   $ ZZ\to H $ &51  &72 & 89 & 96 &  120  \\
    $WW\to HH$ & 0.80 & 1.8   & 3.2 &  4.3&  6.7         \\         
 $ZZ\to HH$ & 0.11 & 0.24   & 0.43 &  0.57&  0.91         \\         
        $WW \to Z H$  & 9.5 & 22       &   33            & 42 & 67  \\   
          $WW\to t\bar t H$  & 0.012 & 0.046     & 0.090    &   0.14 & 0.28 \\
       \hline
    $WW\to Z$  & 2200 &  3100        &  3600         & 4200           &  5200   \\
    $WW\to ZZ$  & 57 & 130       &   200         & 260 & 420 \\   
              \hline
\end{tabular}
\caption{SM Higgs boson production cross sections in units of fb at a muon collider for various energies. For comparison, the SM background processes of $Z$ and $ZZ$ production are also shown.
}
\label{tab:xsecH}
\end{table*}

We will examine the precision measurements of the Higgs boson couplings via the production processes as depicted in Figs.~\ref{fig:VBF} and \ref{fig:Feyn}. 
For instance, at a 10 TeV muon collider with an integrated luminosity of 10 ab$^{-1}$, we may expect the production of about $10^7$ Higgs bosons and $3.6\times 10^4$ Higgs pairs. 
For comparison, we have also included in Table \ref{tab:xsecH} the SM irreducible backgrounds $\mm \stackrel{\rm VBF}{\to} Z, ZZ$, which are also largely from the VBF mechanism, in Table \ref{tab:xsecH}. Although the background rates are larger than the signals by a factor of 4 (55) for the $H$ ($HH$) process, they populate different kinematical regions from the signals and can be reduced by appropriate kinematic cuts. 

\section{$VVH$ Couplings}
\label{sec:VVH}

\begin{figure}[bt]
\centering
\includegraphics[width=0.48\textwidth]{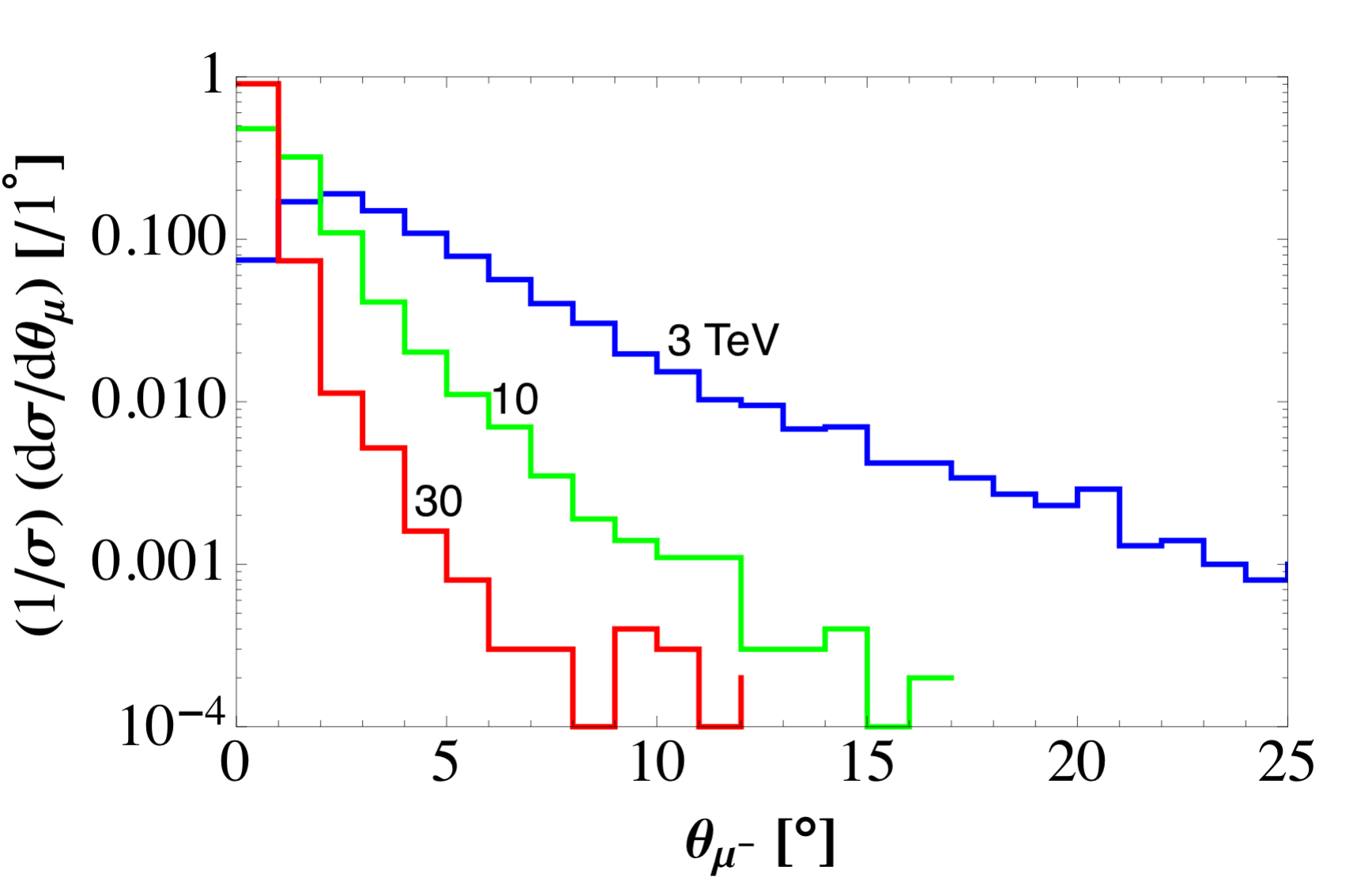}
\includegraphics[width=0.48\textwidth]{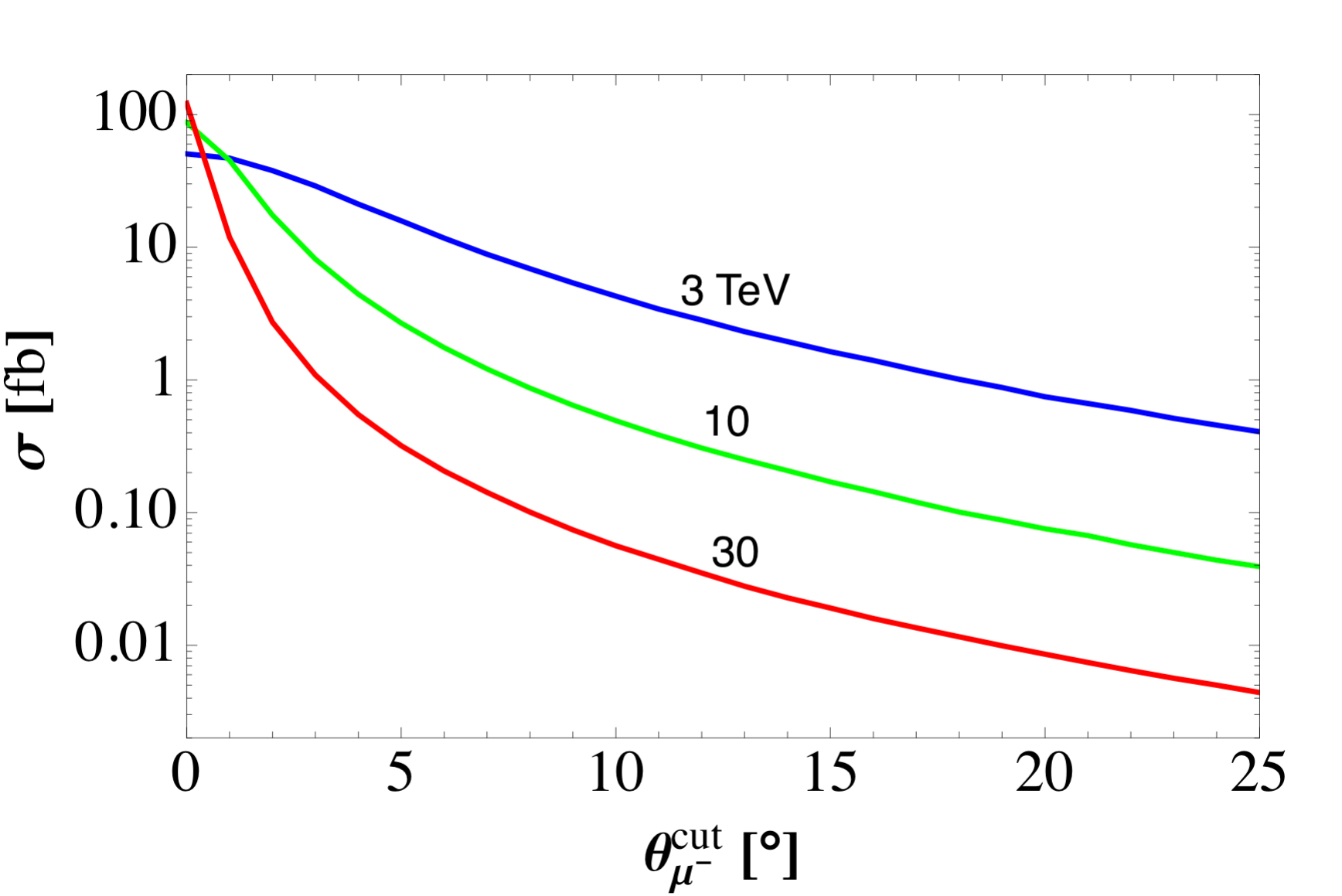}
\caption{$\mm \to \mm H$ via $ZZ$ fusion with $\sqrt s = 3, 10$ and 30 TeV
for (a) angular distribution $\theta_{\mu^-}$, and (b) total cross section versus an angular cut 
$\theta_{\mu^-}^{\rm cut}$. }
\label{fig:thetamuhigh}
\end{figure}

At high energy lepton colliders, the cross section for single $H$ production via the Higgs-strahlung $\mm\to ZH$ falls as $1/s$. The high statistics channels for measurements of $VVH$ couplings rely on the $WW$ and $ZZ$ fusion via the VBF topology:
\bea
\mm &\to& \nu_\mu \bar{\nu}_\mu \ H \qquad \text{($WW$ fusion)} \label{eq:WWfusion}, \\
\mm &\to&  \mm \ H \qquad \text{($ZZ$ fusion)}.  \label{eq:ZZfusion}
\eea
See Fig.~\ref{fig:VBF} for a representative Feynman diagram. 
It would be desirable to separate these two classes of events by tagging the outgoing muons and achieve independent measurements on $WWH$ and $ZZH$ couplings. However, 
for the VBF topology, the outgoing muons have a tendency to stay in the forward region
due to the $t$-channel propagator shown in Fig.~\ref{fig:Feyn}(a). Although the transverse momentum of the outgoing muons is sizable and governed by the propagator mass $p_T^\mu \sim M_Z$, 
at very high energies the muons are all extremely forward with a polar angle typically $\theta_\mu \approx M_Z/E_\mu$. In Fig.~\ref{fig:thetamuhigh}(a), we show the  angular distributions of the outgoing muons at $\sqrt s=3, 10, 30$ TeV. One can see that, for example, the scattering angle for a muon is peaked near $\theta_\mu \sim 0.02\approx 1.2^\circ$ at 10 TeV. These very forward muons would most likely escape the detection in a detector at a few degrees away from colliding beams. 
This feature makes it increasingly difficult to distinguish the processes of the neutral currents 
($ZZ$ fusion \cite{Han:2015ofa}) from the charged currents ($WW$ fusion) at higher energies. 
Therefore, separating these two classes of events would require the capability of detecting very energetic muons in the forward region of a few degrees with respect to the beam. Without this, we would have to focus on
the ``inclusiveness,'' a dominant behavior of the collinear splitting physics recently emphasized in Ref.~\cite{Han:2020uid}.
As a consequence, we will consider two classes of events for VBF production of single $H$:
\begin{itemize}
\item Inclusive channel: events from $WW$ fusion and from $ZZ$ fusion without detecting muons;
\item Exclusive $1\mu$ channel: events from $ZZ$ fusion with at least one muon detected.
\end{itemize}

The inclusive channel is populated predominantly by events from the $WW$ fusion, but also contains  events from $ZZ$ fusion when the outgoing muons go down the beam pipe and escape detection. 
However, as seen from Table \ref{tab:xsecH}, $ZZ$-fusion cross section is roughly 10\% of the $WW$ fusion cross section, and thus a small contamination for the $WWH$ measurement. 
The $1\mu$ channel, on the other hand, comes from the $ZZ$ fusion and is uniquely sensitive to the $ZZH$ coupling, although  it suffers from poor selection efficiency after requiring a muon identification.  In Fig.~\ref{fig:thetamuhigh}(b), we illustrate the  fiducial cross section after the angular acceptance cut $\theta_{\mu^-}^{\rm cut}$. At a fixed angular acceptance, the cross section falls as $\sigma \sim 1/E_\mu^2$. 

\subsection{Inclusive channel}
\label{sec:WWH}

\begin{figure}[tb]
\centering
\begin{subfigure}{0.48\textwidth}
\includegraphics[width=\textwidth]{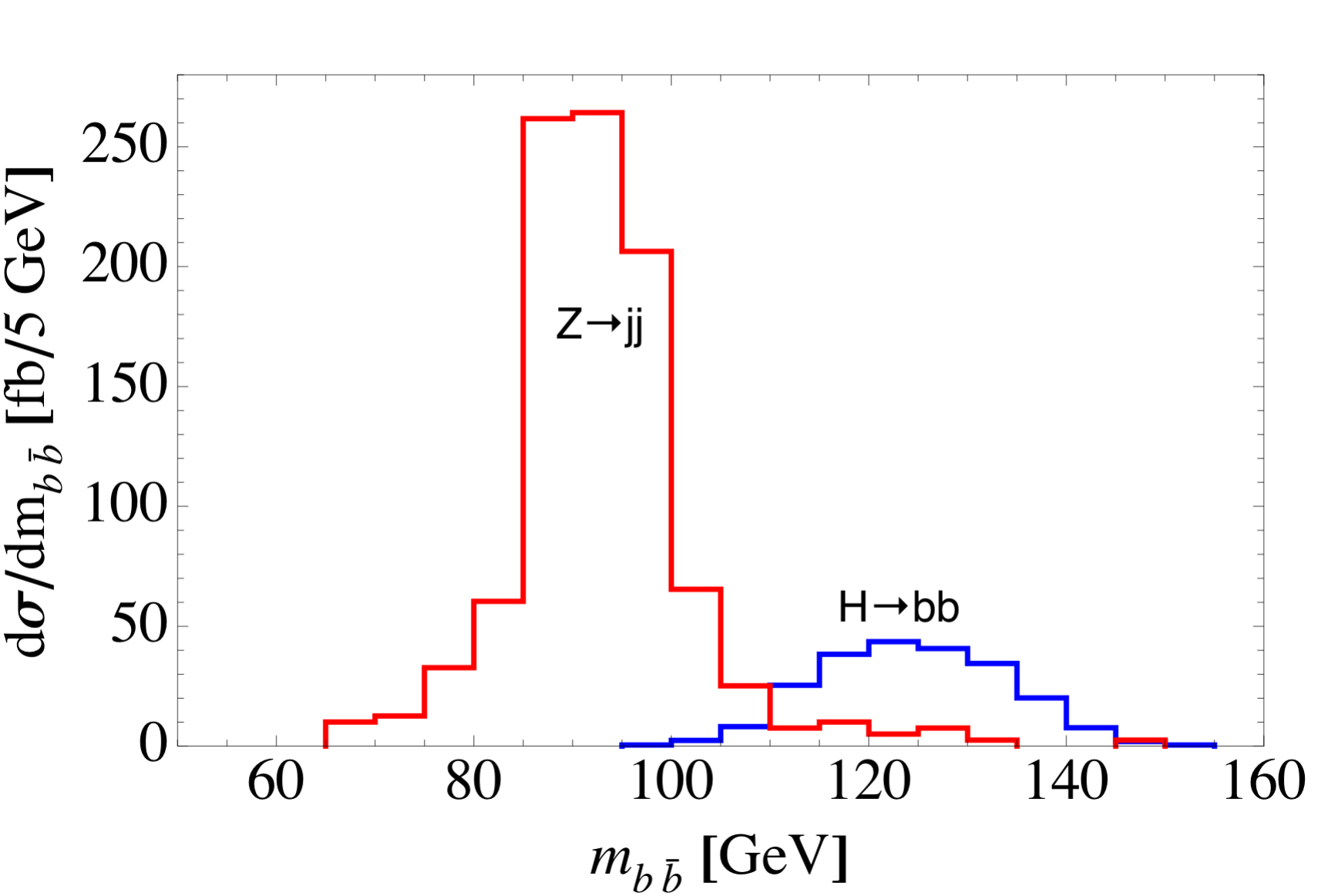}
\caption{}
\end{subfigure}
\begin{subfigure}{0.48\textwidth}
\includegraphics[width=\textwidth]{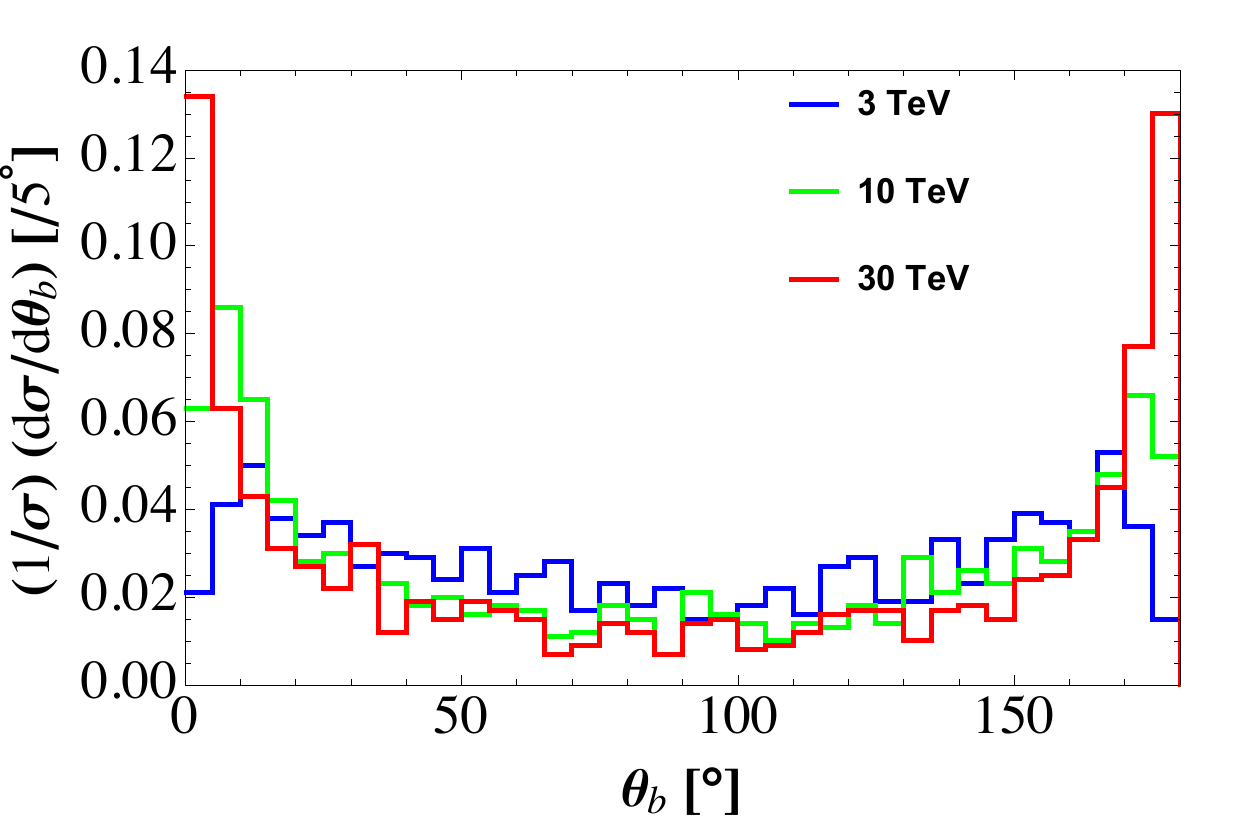}
\caption{}
\end{subfigure}
\caption{(a) Invariant mass distribution for the Higgs boson and $Z$ boson at $\sqrt s=10$ TeV with an energy resolution $10\%$, and (b) the $b$-quark angular distribution $\theta_b$ in the lab frame for $\sqrt s = 3, 10, 30$ TeV.}
\label{fig:mbb}
\end{figure}

Processes contributing to the inclusive channel are shown in Eqs.~(\ref{eq:WWfusion}) and (\ref{eq:ZZfusion}).
We focus on the leading decay channel $H \to b \bar b$. 
The Higgs boson signal will be $b \bar b$ pair near the Higgs mass $m_H$ plus large missing energy, resulting from the missing neutrinos and the undetected muons. 
We impose the basic acceptance cuts on the $b$ jets
\beq
p_T(b) > 30\ {\rm GeV} , \qquad 10^\circ < \theta_b  < 170^\circ ,
\label{eq:cut}
\eeq
where $\theta_b$ is the polar angle of the $b (\bar b)$ jet in the lab frame. 
The irreducible backgrounds, $\mm \to \nu_\mu \bar{\nu}_\mu\ Z$, from either $WW$ fusion shown in Table \ref{tab:xsecH} or $\mm \to ZZ \to \nu_\mu \bar{\nu}_\mu Z$,
which can be readily removed due to the on-shell $Z$ decay $Z\to \nu_\mu \bar{\nu}_\mu$, by a ``recoil mass'' cut
\beq
M_{\rm recoil} = (p_{\mu^+} + p_{\mu^+} - p_H^{})^2 > 200\ {\rm GeV}.
\label{eq:mrec}
\eeq
The key aspect to identify the Higgs signal lies in the resolution to effectively select the $b\bar b$ at the resonant $m_H$. 
In Fig.~\ref{fig:mbb}(a) we  plot the invariant mass distribution for the $H$ signal for $\sqrt s=10$ TeV, after the acceptance cuts and assuming a jet energy resolution of 
\beq
\Delta E/E= 10\%.
\eeq  
For comparison, we have also shown in the same plot the distribution from the $Z$ background. Here we have included all quarks flavors $b,c,s,d,u$. If we demanded a $b$-tagging for our signal selection, we would be able to reduce the $Z\to jj$ background by a factor of 5. However, we do not find the $b$-tagging necessary due to the highly efficient kinematical constraint on $m_{b\bar b}$. In estimating the statistical accuracy for the coupling measurement, we  impose the a mass cut 
\beq
\mbb = m_H\ \pm \ 15\ {\rm GeV} .
\label{eq:mcut}
\eeq
With those cuts, the $Z$ background is essentially removed and we retain the majority of the signal. 
The event selection efficiencies ($\epsilon_{\rm in}$) and the resulting cross sections at different collider energies are summarized in Table \ref{tab:VVHeff} in the top rows.

It is worth noting that, at higher CM energies, the $b$ jets have increasingly small polar angles in the Lab frame and become more forward. The angular distributions for various energies are shown in the right panel of Fig.~\ref{fig:mbb}, where we see the majority of $b$ jets have $\theta_b < 10^\circ$ at $\sqrt{s}=30$ TeV.  This is the reason for the worsening selection efficiencies in Table \ref{tab:VVHeff} as we go to higher CM energies. Obviously, extending the detector angular coverage would significantly increase the signal acceptance. 
If the angular cut on $\theta_b$ in Eq.~(\ref{eq:cut}) is tightened up to $20^\circ - 160^\circ$ instead, the signal reconstruction efficiency will be scaled down by about $10\%$. 

\begin{table*}[tb]
\centering
\begin{tabular}{|c|c|c|c|c|c|}
\hline
  $\sqrt s$ (TeV)   & {3}    & {6} & {10} & {14} & {30} \\
  \hline
  $WW\to H: \ \epsilon_{\rm in}$ (\%) & 54 &  46        &  42         & 39          &  32   \\ 
  $ZZ\to H: \ \epsilon_{\rm in}$ (\%) & 57 &  49        &  44         & 41          &  35   \\ 
  Cross section $\sigma_{\rm in}$ (fb) & 170 &  200   &  220     & 240       &  240   \\            
   \hline
  $ZZ\to H: \ \epsilon_{1\mu}$ (\%) & 11 &  2.7    &  0.84    & 0.37          &  0.071   \\ 
  Cross section $\sigma_{1\mu}$ (fb) & 3.1 &  1.1 & 0.43   &  0.20    &  0.050   \\            
   \hline
  $VV\to HH:\ \epsilon_{hh}$ (\%) & 27 & 18 & 13 & 11 & 7.2   \\ 
  Cross section $\sigma_{hh}$ (ab) & 81 & 140 & 150 & 170 & 200   \\            
   \hline
\end{tabular}
\caption{Selection efficiencies and the estimated cross sections after selection cuts for the inclusive channel, exclusive $1\mu$ channel, as well as the inclusive $HH$ channel.}
\label{tab:VVHeff}
\end{table*}

The total cross section in the inclusive channel can be written, at the leading order, as
\bea
\label{eq:xection0mu}
\sigma_{\rm in} &=& (1+\Delta\kappa_W)^2\ \sigma^{\rm SM}_W + (1+\Delta\kappa_Z)^2\ \sigma^{\rm SM}_Z
\eea
where $\sigma^{\rm SM}_W$ and $\sigma^{\rm SM}_Z$ are the SM cross sections for the $WW/ZZ$ fusion processes. In cases where $\Delta \kappa_{W/Z} \ll 1$, the linear terms dominate which, in the EFT language, is equivalent to keeping only the interference term from the dim-6 operators. We do not make such an assumption in the $\kappa$-scheme adopted in this work.

In this subsection we will vary $\kw$ and $\kz$ one at a time, and consider a simultaneous fit to both parameters later in this section. The 95\% confidence level (C.L.) sensitivities in the relative errors $\Delta\kappa_{W/Z}$ are shown in Table \ref{tab:VVHinfit}. 
The achievable accuracies are impressive, comparing with the anticipated best results $\Delta\kw \sim 0.6\%$ from the ILC/CLIC and $\Delta\kz \sim 0.2\%$ from the expectations at the Higgs factories \cite{Abada:2019zxq,An_2019}. 

\begin{table*}[tb]
\centering
\begin{tabular}{|c|c|c|c|c|c|}
\hline
  $\sqrt s$ (TeV)   & {3}    & {6} & {10} & {14} & {30} \\
  benchmark lumi (ab$^{-1}$)   & {1}    & {4} & {10} & {20} & {90} \\
  \hline
  ($\Delta \kw)_{\rm in}$ & 0.26\% &  0.12\%        &  0.073\%         & 0.050\%          &  0.023\%   \\ 
   ($\Delta \kz)_{\rm in}$ & 2.4\% &  1.1\%       &  0.65\%         & 0.46\%          &  0.20\%   \\   
   \hline
      ($\Delta \kz)_{1\mu}$ & 1.7\% &  1.5\%       &  1.5\%         & 1.5\%          &  1.5\%   \\    
   \hline
\end{tabular}
\caption{The 95\% C.L.~in $\Delta\kappa_{W/Z}$ in the inclusive channel by varying one coupling at a time, as well as for $\Delta\kz$ from the exclusive $1\mu$ process.} 
\label{tab:VVHinfit}
\end{table*}

\subsection{Exclusive $1\mu$ channel}
\label{sec:ZZH}

The leading process contributing to the exclusive $1\mu$ channel is $ZZ$ fusion in Eq.~(\ref{eq:ZZfusion}), whose rate is shown in Table \ref{tab:xsecH}. 
Again, with the same decay mode, the Higgs boson signal will be $b \bar b$ pair near the Higgs mass $m_H$ plus $\mm$ in the forward-backward regions. 
The leading background is $\mm \to ZZ \to \mm Z$ with $Z\to b \bar b$. There is no $WW$ fusion analogue for this channel. We adopt the same basic cuts as in Eqs.~(\ref{eq:cut}), (\ref{eq:mrec}) and (\ref{eq:mcut}). The background is highly suppressed.  In addition, we require the presence of at least one muon to be in
\beq
\quad 10^\circ < \theta_{\mu^\pm}  < 170^\circ .
\label{eq:mucut}
\eeq
This turns out to be very costly to the signal, since the majority of the muons have $\theta_\mu < 10^\circ$, as already seen in Fig.~\ref{fig:thetamuhigh}. As such, the signal reconstruction efficiencies for this channel are very low and are shown in Table \ref{tab:VVHeff}, together with the predicted cross sections in the middle rows. 
With the high luminosity expected, the 95\% C.L. on the coupling measurements is shown also in Table \ref{tab:VVHinfit} for the exclusive $1\mu$ channel.
Although the result at a 3 TeV collider is comparable to that from the inclusive channel, at higher energies the estimated precision is worse than  the inclusive channel despite the higher energies and more luminosities. This is mainly due to the significantly reduced number of events from the tagging requirement for a forward-backward muon.  

\begin{figure}[tb]
\centering
\begin{subfigure}{0.48\textwidth}
\includegraphics[width=\textwidth]{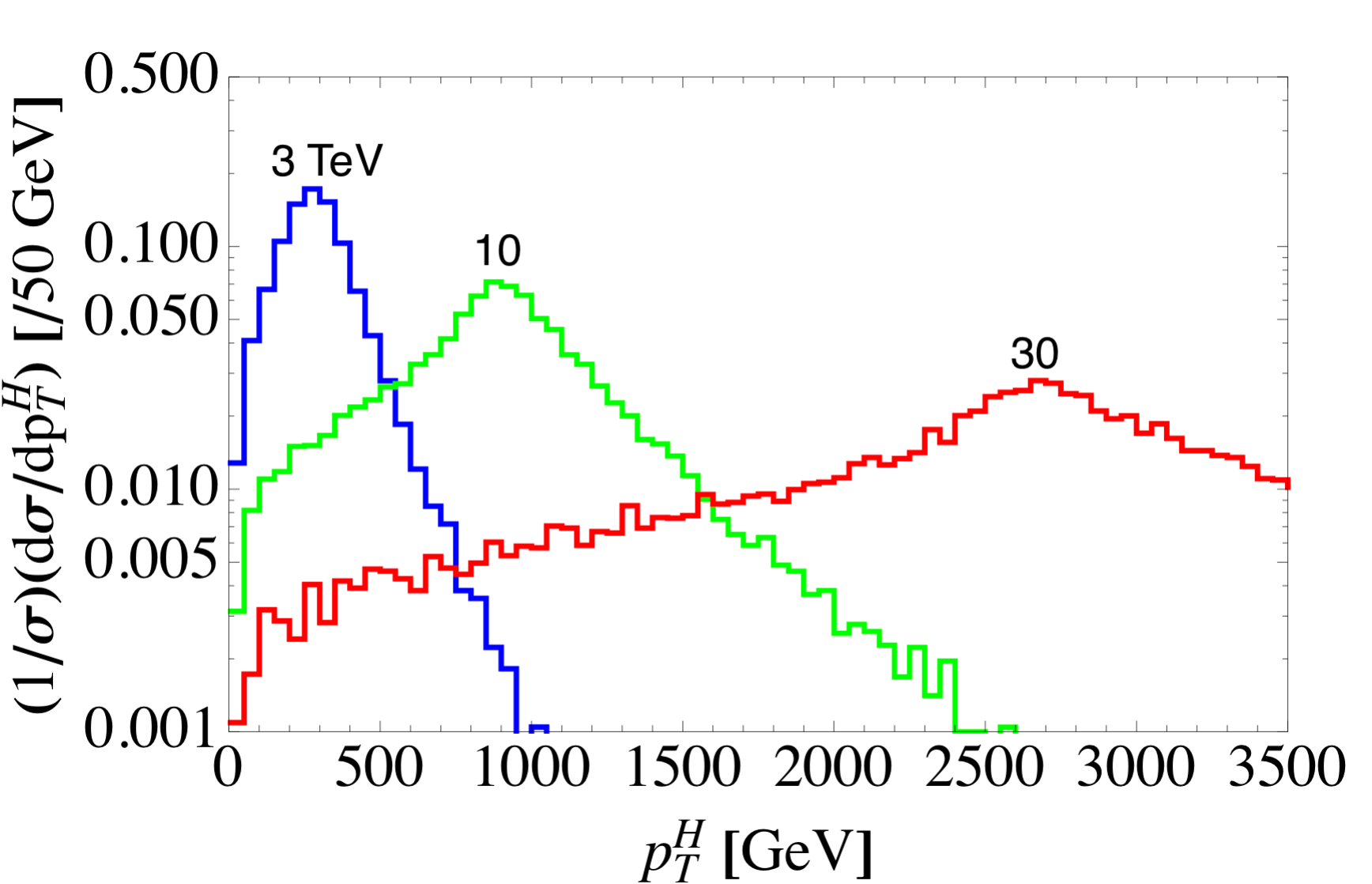}
\caption{}
\end{subfigure}
\begin{subfigure}{0.48\textwidth}
\includegraphics[width=\textwidth]{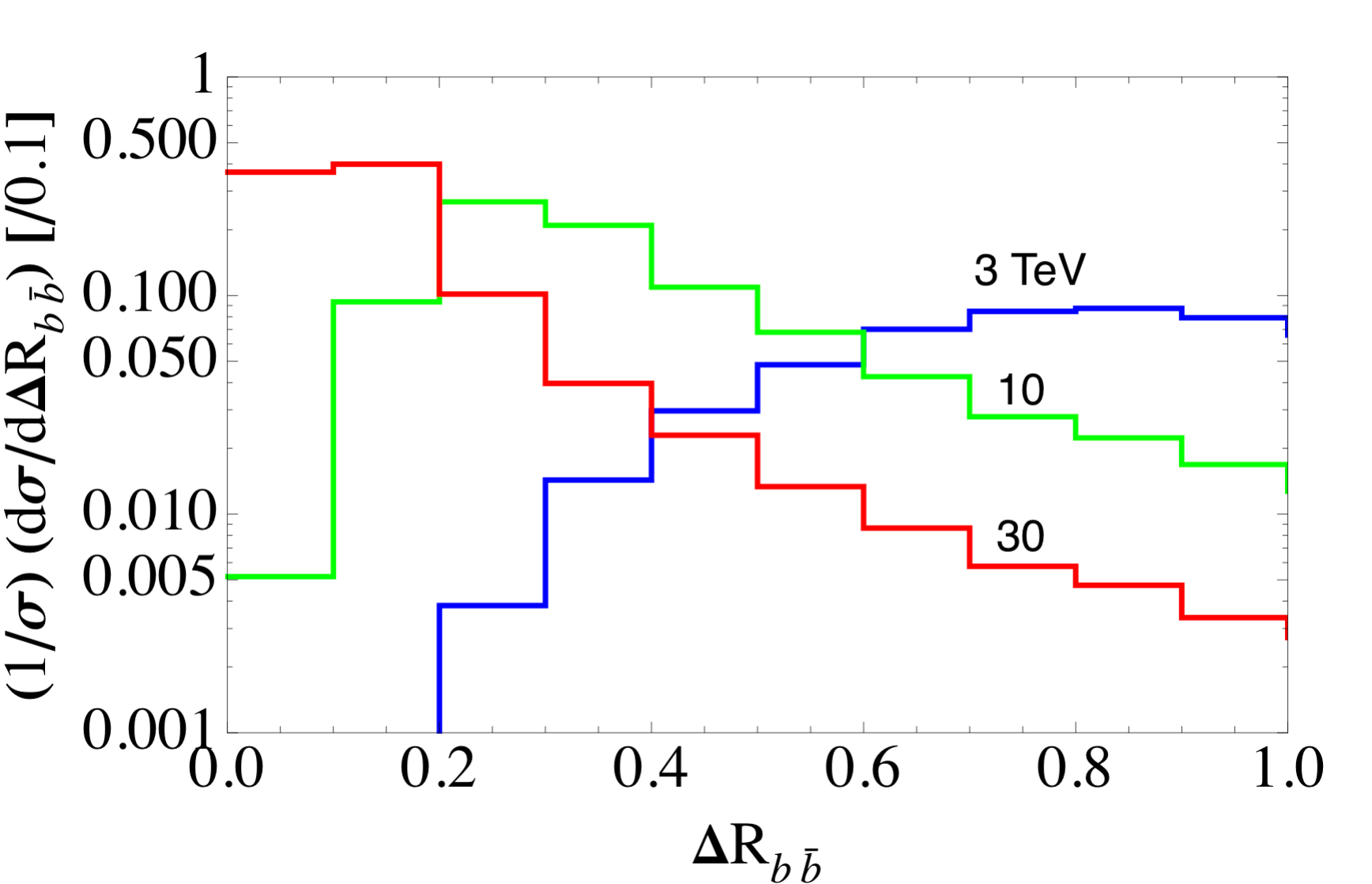}
\caption{}
\end{subfigure}
\caption{(a) $p_T^{H}$ distribution of the Higgs boson in $1\mu$ channel (b) Separation of the $b$ jets from $H\to b\bar{b}$. }
\label{fig:boostedbb}
\end{figure}

It is important to note another significant consequence of requiring one muon in the range of $\quad 10^\circ < \theta_{\mu^\pm}  < 170^\circ$. For highly energetic muons, this large scattering angle leads to a high transverse momentum $p_T^\mu > 0.17E_\mu $ and, consequently, induces a strong recoil in the Higgs boson produced in the final state. In Fig.~\ref{fig:boostedbb} we show the $p_T$ distribution of the Higgs boson in (a) for the $1\mu$ channel as well as $R_{bb}$ in (b), the separation of the $b$-jets from $H\to b\bar{b}$. In particular, at $\sqrt{s}=30$ TeV, the Higgs boson tend to have a large $p_T$, in the order of 2.5 TeV, and the resulting decay is boosted with $R_{bb}\sim 0.2$. Care needs to be taken when reconstructing such  boosted events.

\subsection{Two-parameter likelihood fit of $\kappa_W$ and $\kappa_Z$}

\begin{figure}[tb]
\centering
\centering
\begin{subfigure}{0.45\textwidth}
\includegraphics[width=\textwidth]{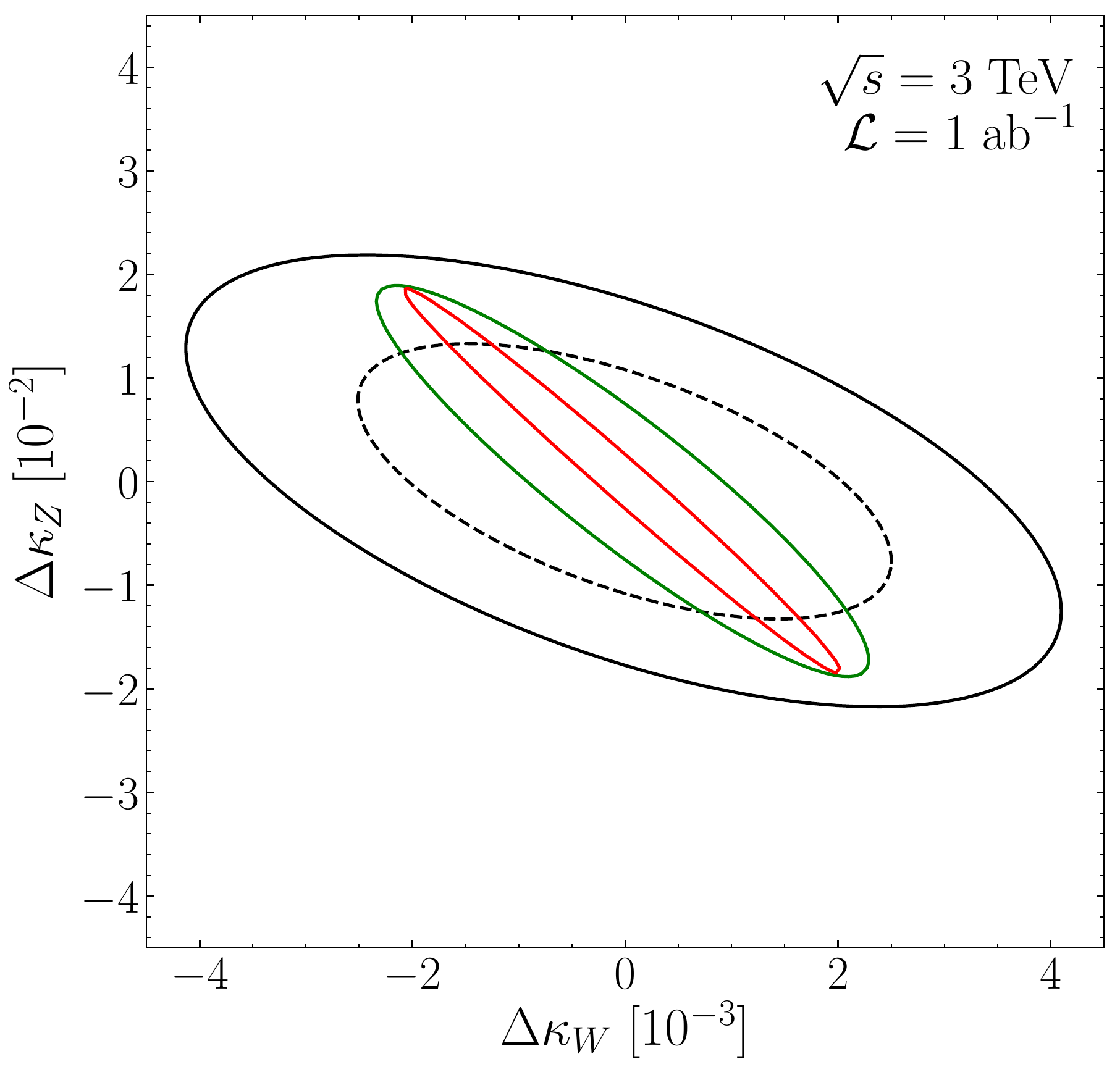}
\caption{}
\end{subfigure}
\begin{subfigure}{0.45\textwidth}
\includegraphics[width=\textwidth]{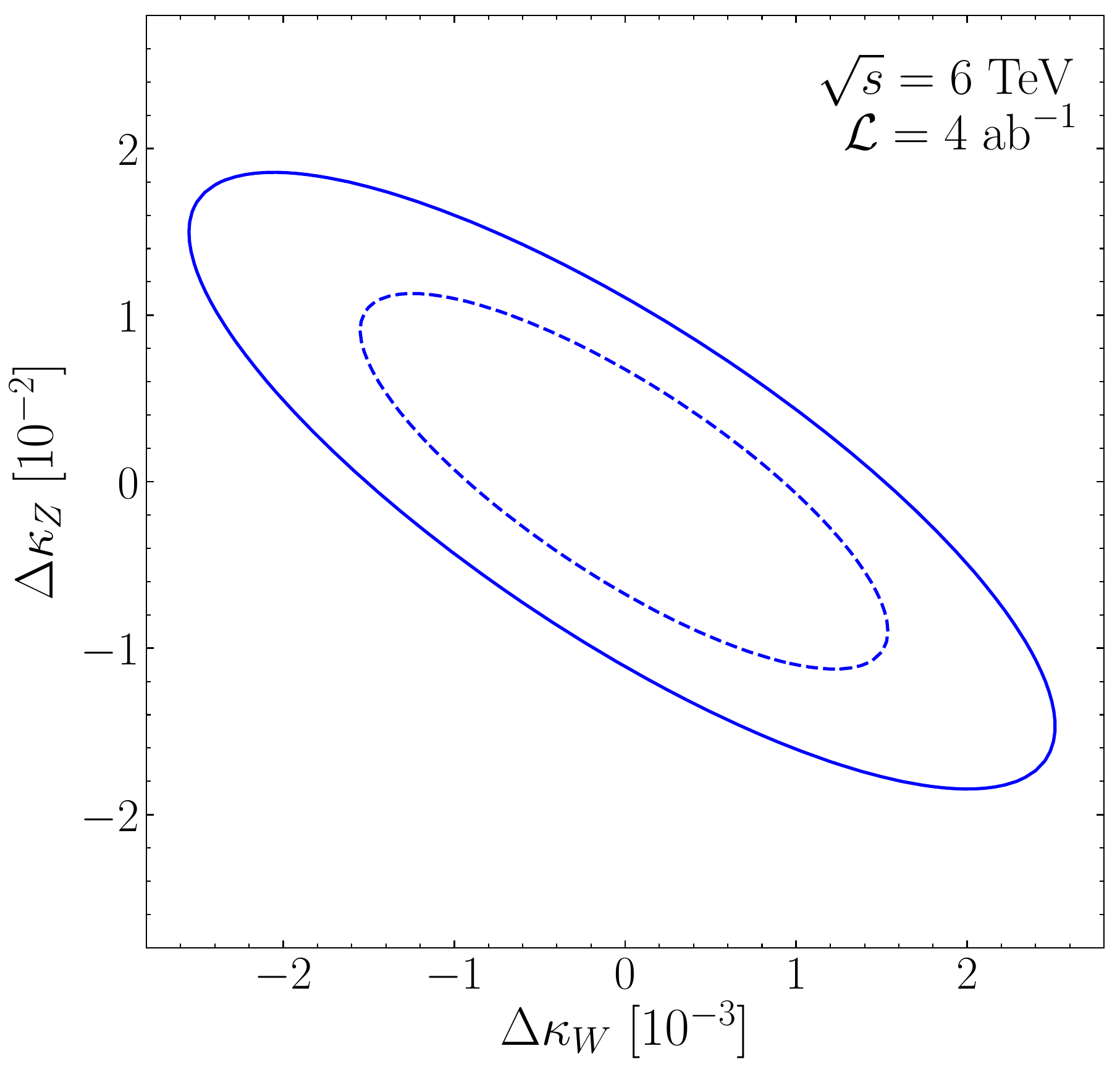}
\caption{}
\end{subfigure}\\
\begin{subfigure}{0.45\textwidth}
\includegraphics[width=\textwidth]{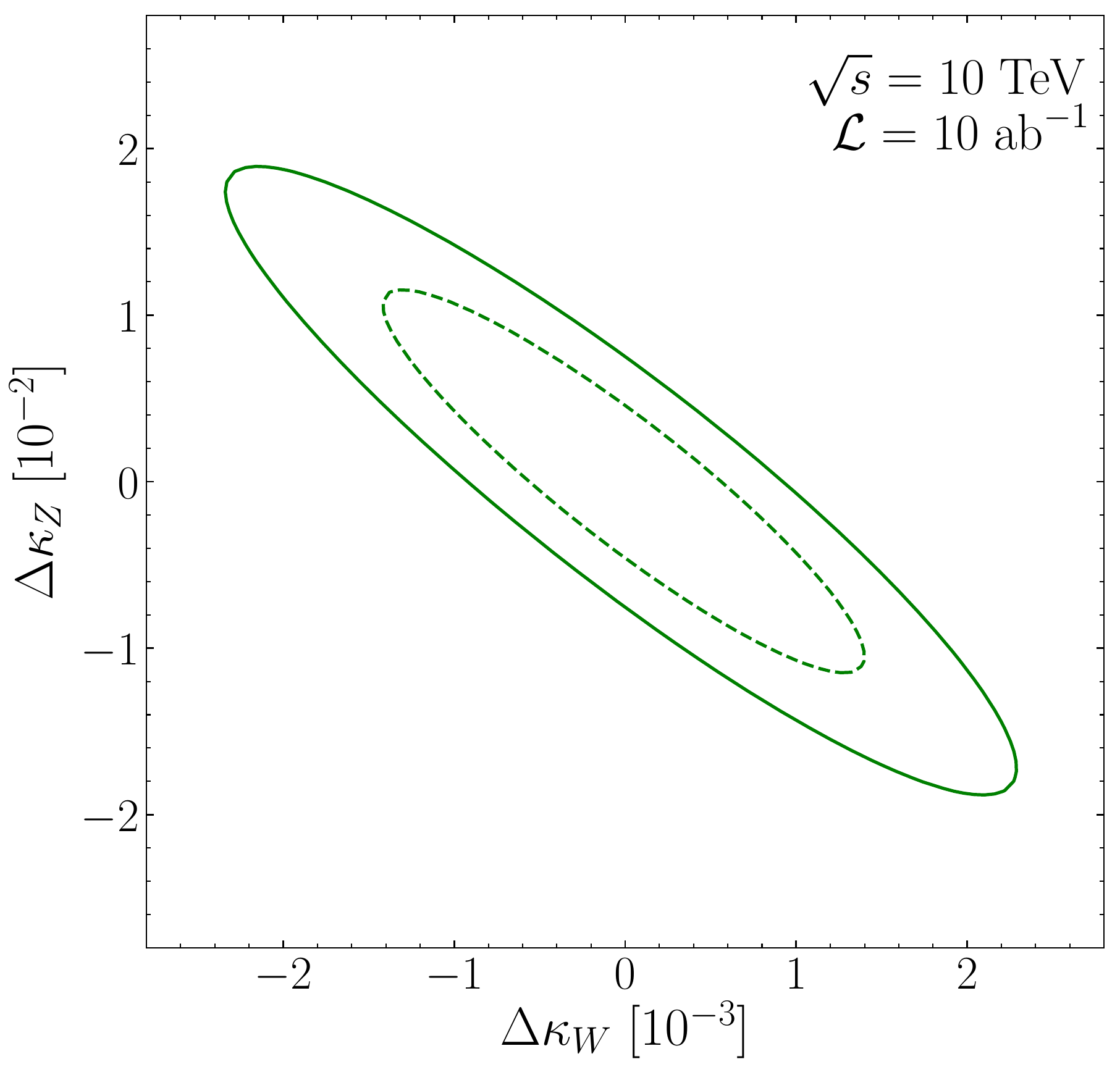}
\caption{}
\end{subfigure}
\begin{subfigure}{0.45\textwidth}
\includegraphics[width=\textwidth]{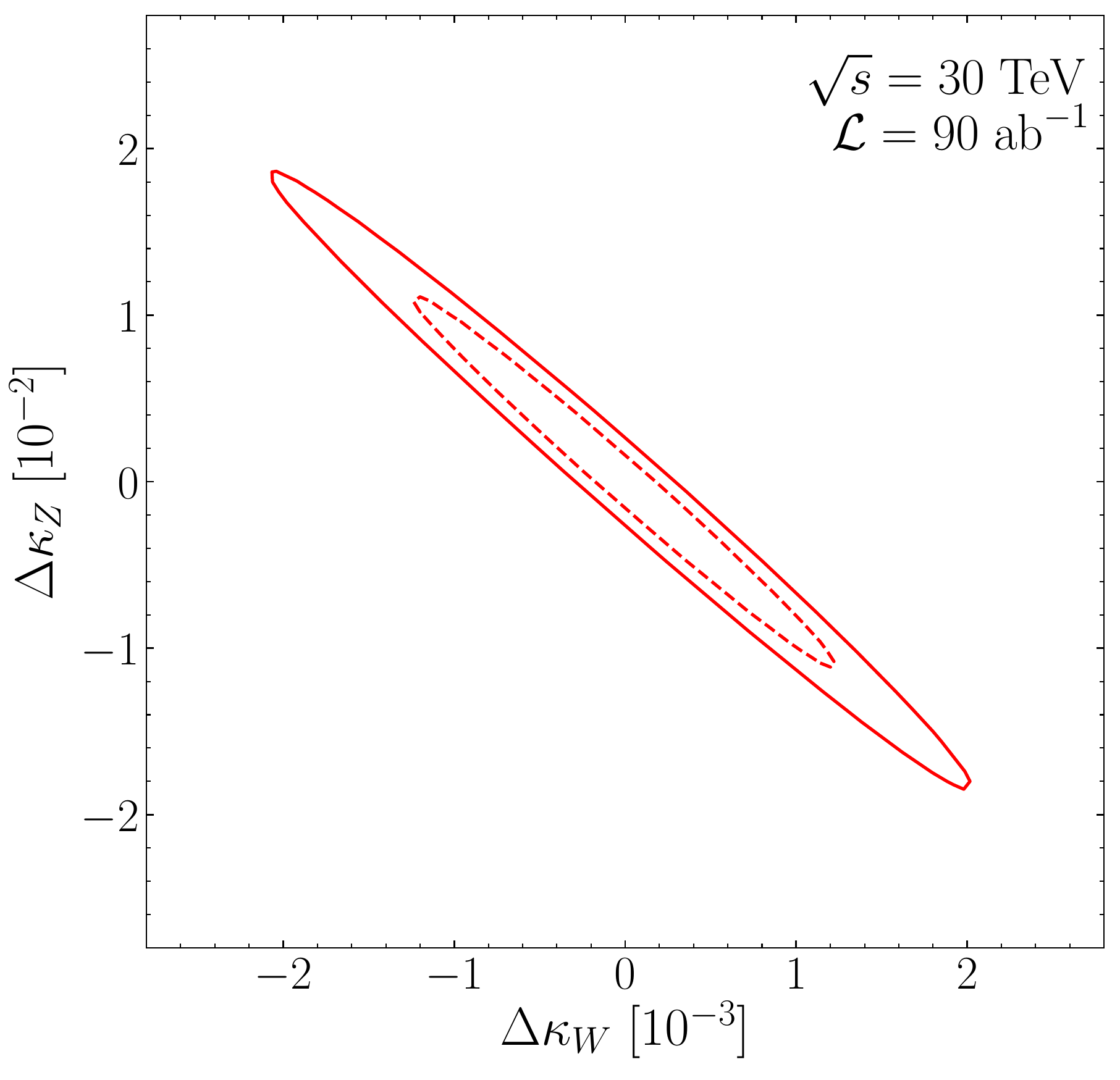}
\caption{}
\end{subfigure}
\caption{Correlated bounds with 95\% C.L.~(solid) and 68\% C.L.~(dashed) in the $\Delta\kappa_W$-$\Delta\kappa_Z$ plane for $\sqrt s=3,6,10,30$ TeV, respectively. In (a), inner ellipses (solid) include the 95\% C.L.~results for 10 TeV and 30 TeV for comparison. 
}
\label{fig:wz}
\end{figure}

In this subsection we perform a two-bin likelihood fit of $\kappa_W$ and $\kappa_Z$ making use of  the inclusive and exclusive $1\mu$ channels. 
We construct a Poisson log-likelihood function 
\begin{equation}
{\rm LL} = \ln\frac{e^{-N(\Delta\kappa_W, \Delta\kappa_{Z})}[N(\Delta\kappa_W, \Delta\kappa_{Z})]^{N_{\rm SM}}}{N_{\rm SM}!},
\end{equation}
where the numbers of events are
\begin{equation}
N(\Delta\kappa_W, \Delta\kappa_{Z}) = \sigma(\Delta\kappa_W,\Delta\kappa_{Z}) \mathcal{L}_{\rm lumi}\ ,\quad N_{\rm SM} = \sigma(\Delta\kappa_W =0, \Delta\kappa_{Z} =0) \mathcal{L}_{\rm lumi}\ ,
\end{equation}
and $\mathcal{L}_{\rm lumi}$ is the integrated luminosity.
We compute such likelihood function for each channel and a global likelihood as the product of the individual ones. Then we compute the 68\% and 95\% C.L. regions on the 
$\Delta\kw$-$\Delta\kz$ plane, corresponding to ${\rm LL} = {\rm LL}_{\rm max} - 1.15$ and ${\rm LL} = {\rm LL}_{\rm max} - 3.10$, respectively. The resulting contours are shown in Fig.~\ref{fig:wz}.
As expected, the precision for $\Delta\kw$ is better than $\Delta\kz$ by about an order of magnitude at high energies. The projection of the elipses onto the $\Delta \kappa_W$-axis in Fig.~\ref{fig:wz} gives the uncertainty marginalized over $\Delta \kappa_Z$, and vice versa. The resulting errors are larger than those in the single parameter fit, which varies one parameter at a time and assumes SM values for the rest.

\section{$HHH$ and $WWHH$ Couplings}
\label{sec:HH}

Pair production of the Higgs boson provides a direct measurement on the trilinear $HHH$ and quartic $VVHH$ couplings. The main advantage of a high-energy collider, with $\sqrt{s}\gg2 m_H$, lies in the capability to copiously produce Higgs boson pairs. At a high-energy muon collider, as shown in Sec.~\ref{sec:Higgs}, one would expect about 36,000 (68,000) $HH$ at 10 TeV (30 TeV).
To probe the Higgs self-coupling, we utilize the VBF mechanism for the inclusive double Higgs production 
\begin{equation}
\label{eq:HHinclu}
\mu^+\mu^-\ \stackrel{VBF}\longrightarrow\ HH + X\ ,
\end{equation}
where $X=\nu\bar{\nu}$ for $WW$ fusion and $\mu^+\mu^-$ for the $ZZ$ fusion. As can be seen from the Feynman diagrams in Fig.~\ref{fig:Feyn}, the $HH$ production involves three classes of couplings: $\kappa_W, \kappa_3$ and $\kappa_{W2}$. Since $\kappa_W$ can be measured very precisely from the single Higgs production, as shown in Section \ref{sec:VVH}, we will assume in the current section that $\kappa_W=1$ as in the SM and study the interplay of $\kappa_3$ and $\kappa_{W2}$ in the $HH$ production. As discussed in Section \ref{sec:VVH}, the outgoing remnant particles tend to stay in the forward region and escape detection. Therefore, similar to the single Higgs production, we will consider the inclusive channel in Eq.~(\ref{eq:HHinclu}), which is populated dominantly by the $WW$ fusion and, to a less extent, by the $ZZ$ fusion events when the outgoing muons are too forward to be detected.

The cross section for the inclusive $\mu^+\mu^-\rightarrow  HH+X$ can be parametrized as~\cite{Contino:2013gna} 
\begin{equation}
\sigma = \sigma_{\rm SM}\left[1 + R_1 \Delta\kappa_{W_2} + R_2\Delta\kappa_3 + R_3 \Delta\kappa_{W_2}\Delta\kappa_3 + R_4 \left(\Delta\kappa_{W_2}\right)^2 + R_5 \left(\Delta\kappa_3\right)^2\right],
\label{eq:HH_xsec_1}
\end{equation}
where the $\sigma_{\rm SM}$ is the SM cross section. The SM cross section $\sigma_{\rm SM}$ and coefficients $R_i$, before any cuts, are given in Table~\ref{tab:hh_sec_1}.
\begin{table}
\centering
\begin{tabular}{ccccccc}
\hline
  $\sqrt{s}$ [TeV]  &  $\sigma_{\rm SM}$ [fb]  &  $R_1$  &  $R_2$  &  $R_3$  &   $R_4$   &  $R_5$  \\
\hline
       3 TeV        &          $0.91$          & $-3.5$  & $-0.65$ &  $3.1$  &   $14$    & $0.49$  \\
       6 TeV        &           $2.0$            & $-3.9$  & $-0.50$  &  $2.8$  &   $29$    & $0.35$  \\
       10 TeV       &          $3.6$           & $-4.3$  & $-0.43$ &  $2.7$  &   $54$    & $0.29$  \\
       14 TeV       &          $4.9$           & $-4.4$  & $-0.38$ &  $2.6$  &   $80$    & $0.25$  \\
       30 TeV       &          $7.6$           & $-4.4$  & $-0.28$ &  $2.3$  & $210$ & $0.19$  \\
\hline
\end{tabular}
\caption{Predicted cross sections of the inclusive $\mu^+\mu^-\rightarrow  HH+X$, as given in Eq.~(\ref{eq:HH_xsec_1}) at different muon collider energies.}
\label{tab:hh_sec_1}
\end{table}
It is instructive to consider the energy dependence of different classes of Feynman diagrams contributing to $HH$ production, by studying the partonic scattering $W^+W^-\rightarrow HH$. As the dominant contribution comes from the longitudinal $W$ scattering $W^+_LW^-_L\rightarrow HH$, the scattering amplitude can be written as
\begin{equation}
\mathcal{A}(W^+_LW^-_L\rightarrow HH) = \mathcal{A}_{\rm SM} + \mathcal{A}_1 \Delta\kappa_{W_2} + \mathcal{A}_2 \Delta\kappa_3,
\end{equation}
where $\mathcal{A}_{\rm SM},\ \mathcal{A}_2 \sim$ constant, and $\mathcal{A}_1 \sim E^2$ at high energies $E\gg M_W$. Because of the energy growing behavior of $\mathcal{A}_1$, the cross section has a strong dependence on $\Delta\kappa_{W_2}$ over a large range of phase space. As a result, we expect to be able to constrain $\kappa_{W_2}$ better than $\kappa_3$. This argument also shows, when extracting the trilinear Higgs self-coupling it is important to consider the impact from the quartic $VVHH$ coupling.  In this study, we have assumed the $HHVV$ vertex is modified only in its strength for simplicity, while in many well-motivated new physics models the tensor structure of the quartic coupling could also be corrected \cite{Liu:2018vel,Liu:2018qtb}. It will be interesting to further assess the impact of these additional modifications on the extraction of $\kappa_3$ \cite{VVHHfurther}. 

For the Higgs decays, we once again focus on the leading decay channel $HH \to b \bar b\ b \bar b$, which has a SM branching fraction ${\rm BR}(4b)\simeq 34\%$. We impose basic acceptance cuts
\begin{equation}
p_T(b) > 30~{\rm GeV},\quad 10^\circ < \theta_b < 170^\circ,\quad \Delta R_{bb} > 0.4.
\end{equation}
As before, we further assume the jet energy resolution to be $\Delta E/E = 10\%$.

The Higgs candidates are reconstructed from the four most energetic jets. The four jets are paired by minimizing
\begin{equation}
(m_{j_1j_2} - m_H)^2 + (m_{j_3j_4} - m_H)^2.
\end{equation}
And for each Higgs candidate, we impose 
\begin{equation}
|m_{jj} - m_H| < 15~{\rm GeV}
\end{equation}
to reject background from $Z$ and $W$ resonances. We also require the recoil mass
\begin{equation}
M_{\rm recoil} = \sqrt{(p_{\mu^+} + p_{\mu^-} - p_{H_1} - p_{H_2})^2} > 200~{\rm GeV}.
\end{equation}
The signal selection efficiencies and the corresponding cross sections are listed in Table \ref{tab:VVHeff}.
If we tighten the angular cut to $20^\circ$, the efficiencies would drop by a factor of $3$ -- $4$. 

\begin{table}
\centering
\begin{tabular}{ccccccc}
\hline
  $m_{HH}$ [GeV]  &  $\sigma_{\rm SM}$ [ab]  &  $r_1$  &  $r_2$  &  $r_3$  &   $r_4$   &  $r_5$  \\
\hline
    $[0, 350)$    &           $15$           & $-2.7$  & $-1.7$  &  $7.6$  &   $6.7$   &  $2.6$  \\
   $[350, 450)$   &           $24$           & $-3.4$  & $-1.2$  &  $5.2$  &   $7.8$   & $0.95$  \\
   $[450, 550)$   &           $24$           &  $-4.0$   & $-0.91$ &  $4.6$  &   $12$    & $0.52$  \\
   $[550, 650)$   &           $21$           & $-4.6$  & $-0.70$  &  $4.7$  &   $17$    & $0.36$  \\
   $[650, 750)$   &           $17$           & $-5.3$  & $-0.60$  &  $5.1$  &   $26$    & $0.28$  \\
   $[750, 950)$   &           $24$           & $-6.9$  & $-0.52$ &  $6.3$  &   $46$    & $0.23$  \\
  $[950, 1350)$   &           $23$           &  $-11$  & $-0.47$ &  $8.7$  & $120$ & $0.19$  \\
  $[1350, 5000)$  &           $15$           &  $-18$  & $-0.30$  &  $7.2$  & $240$ & $0.075$ \\
\hline
\end{tabular}
\caption{Cross sections of the inclusive $\mu^+\mu^-\rightarrow  HH + X \rightarrow b\bar{b}\ b\bar{b} + X$ in different $m_{HH}$ ranges as the coefficients corresponding to the five terms in Eq.~(\ref{eq:rs}) with $\sqrt s =10$ TeV.}
\label{tab:rs10TeV}.
\end{table}

\begin{table*}[tb]
\centering
\begin{tabular}{|c|c|c|c|c|c|}
\hline
  $\sqrt s$ (TeV)   & {3}    & {6} & {10} & {14} & {30} \\
   benchmark lumi (ab$^{-1}$)   & {1}    & {4} & {10} & {20} & {90} \\
  \hline
  ($\Delta \kappa_{W_2})_{\rm in}$ & 5.3\% &  1.3\%        &  0.62\%         & 0.41\%      &  0.20\%   \\ 
   ($\Delta \kappa_3)_{\rm in}$ & 25\% &  10\%        &  5.6\%         & 3.9\%      &  2.0\%   \\   
   \hline
\end{tabular}
\caption{The 95\% C.L.~in $\Delta \kappa_{W_2}$ and $\Delta \kappa_3$ for the inclusive channel, by varying one coupling at a time.} 
\label{tab:VVHHfit}
\end{table*}

\begin{figure}[tb]
\centering
\begin{subfigure}{0.45\textwidth}
\includegraphics[width=\textwidth]{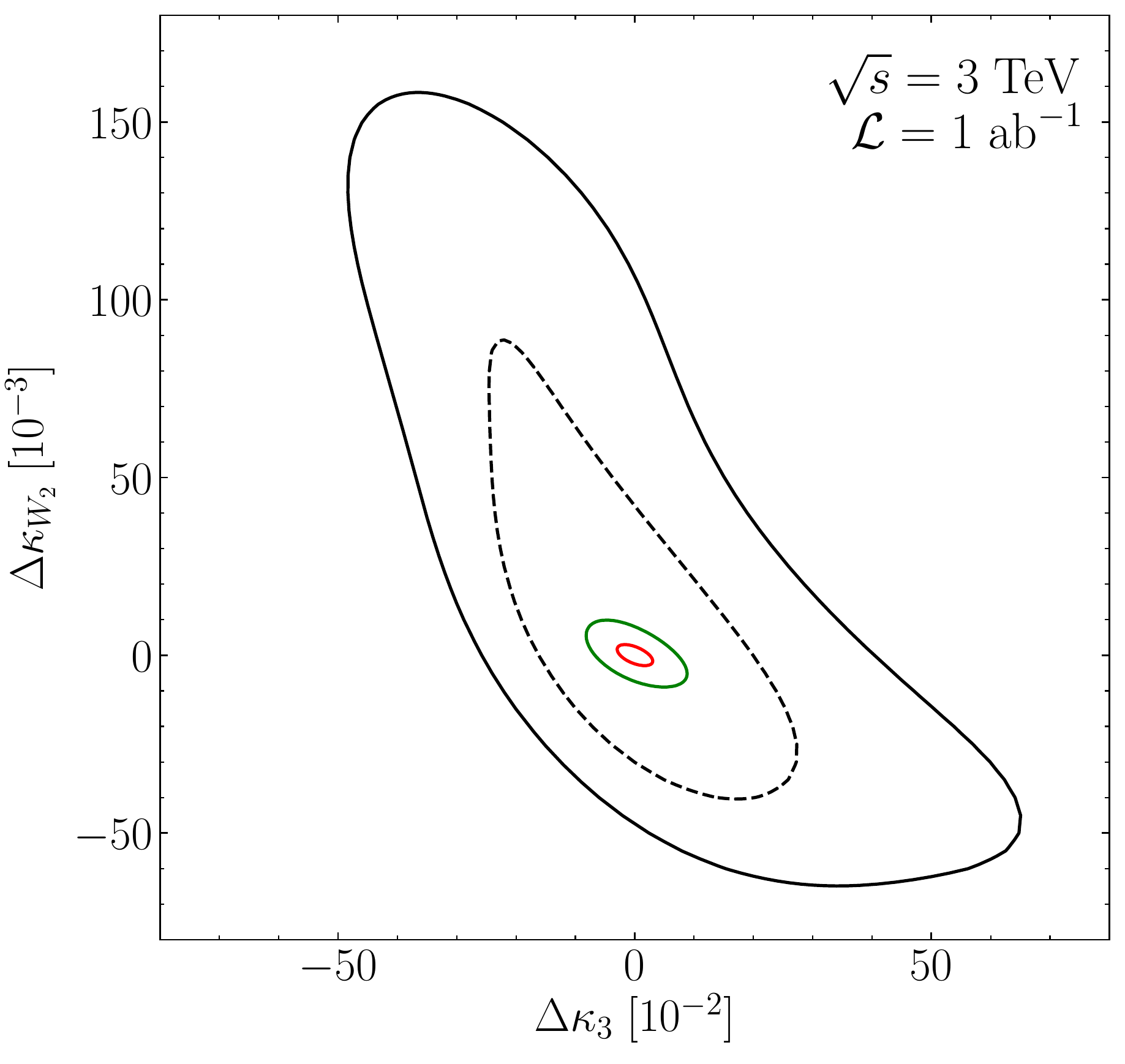}
\caption{}
\end{subfigure}
\begin{subfigure}{0.45\textwidth}
\includegraphics[width=\textwidth]{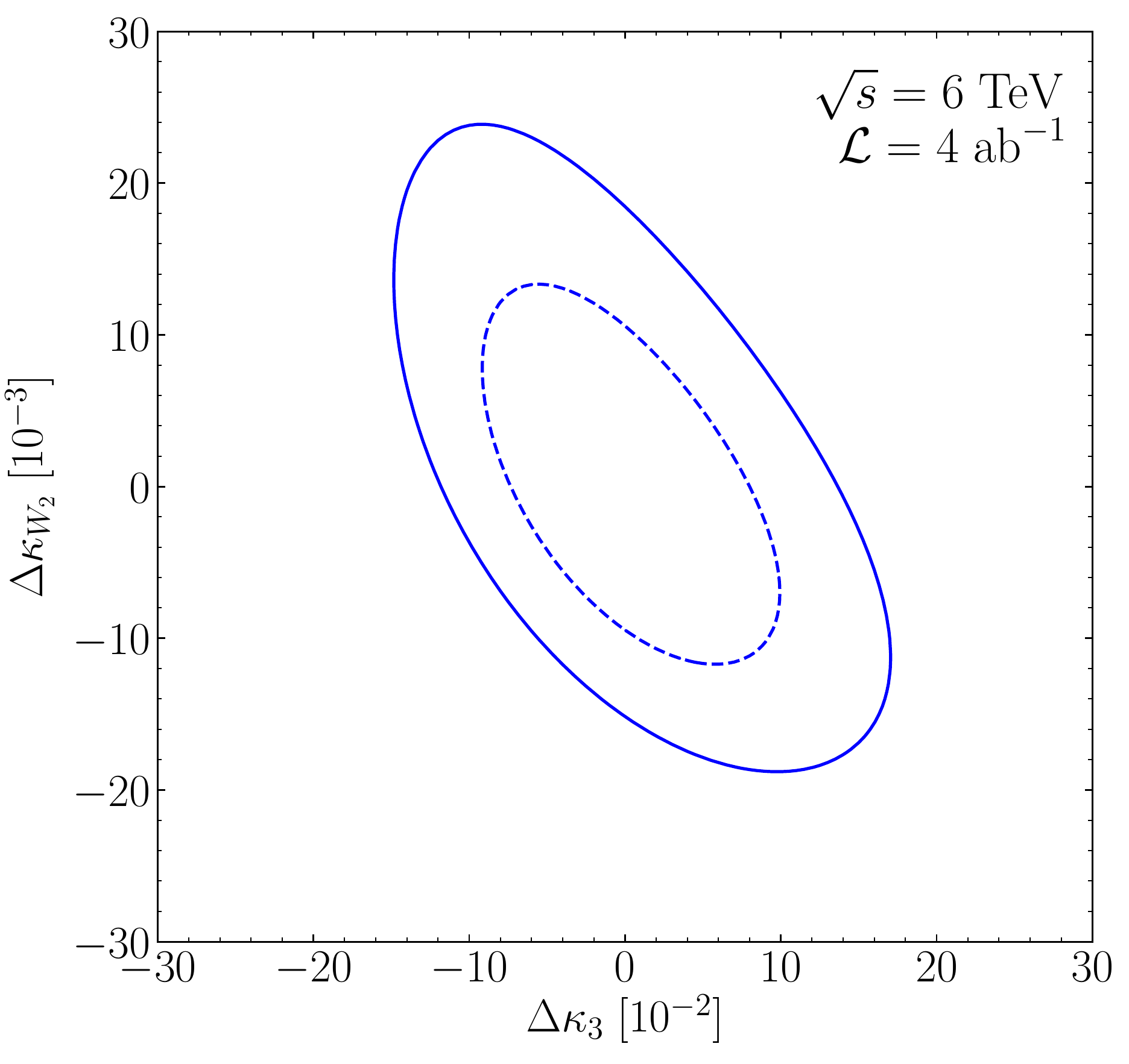}
\caption{}
\end{subfigure}\\
\begin{subfigure}{0.45\textwidth}
\includegraphics[width=\textwidth]{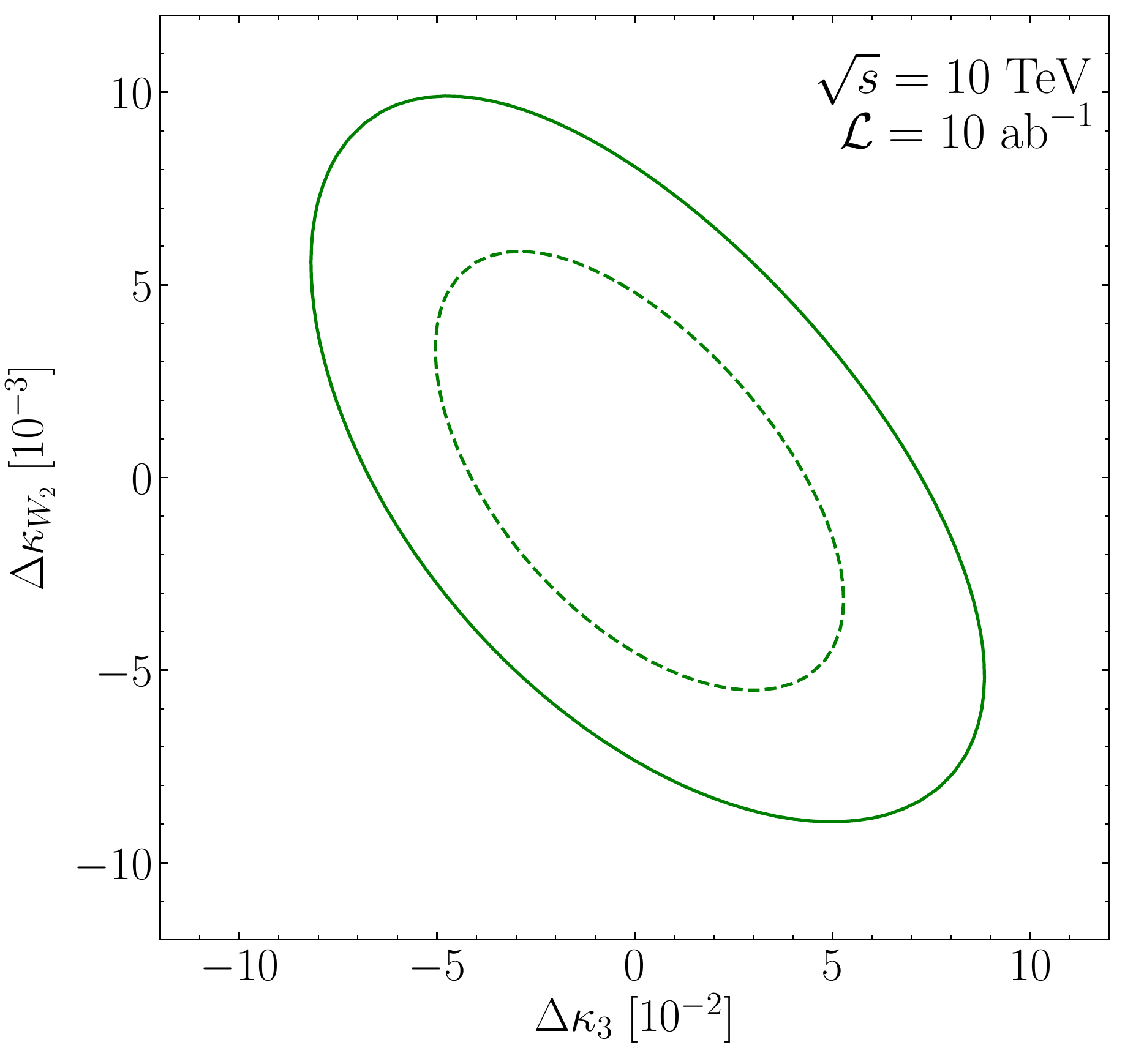}
\caption{}
\end{subfigure}
\begin{subfigure}{0.45\textwidth}
\includegraphics[width=\textwidth]{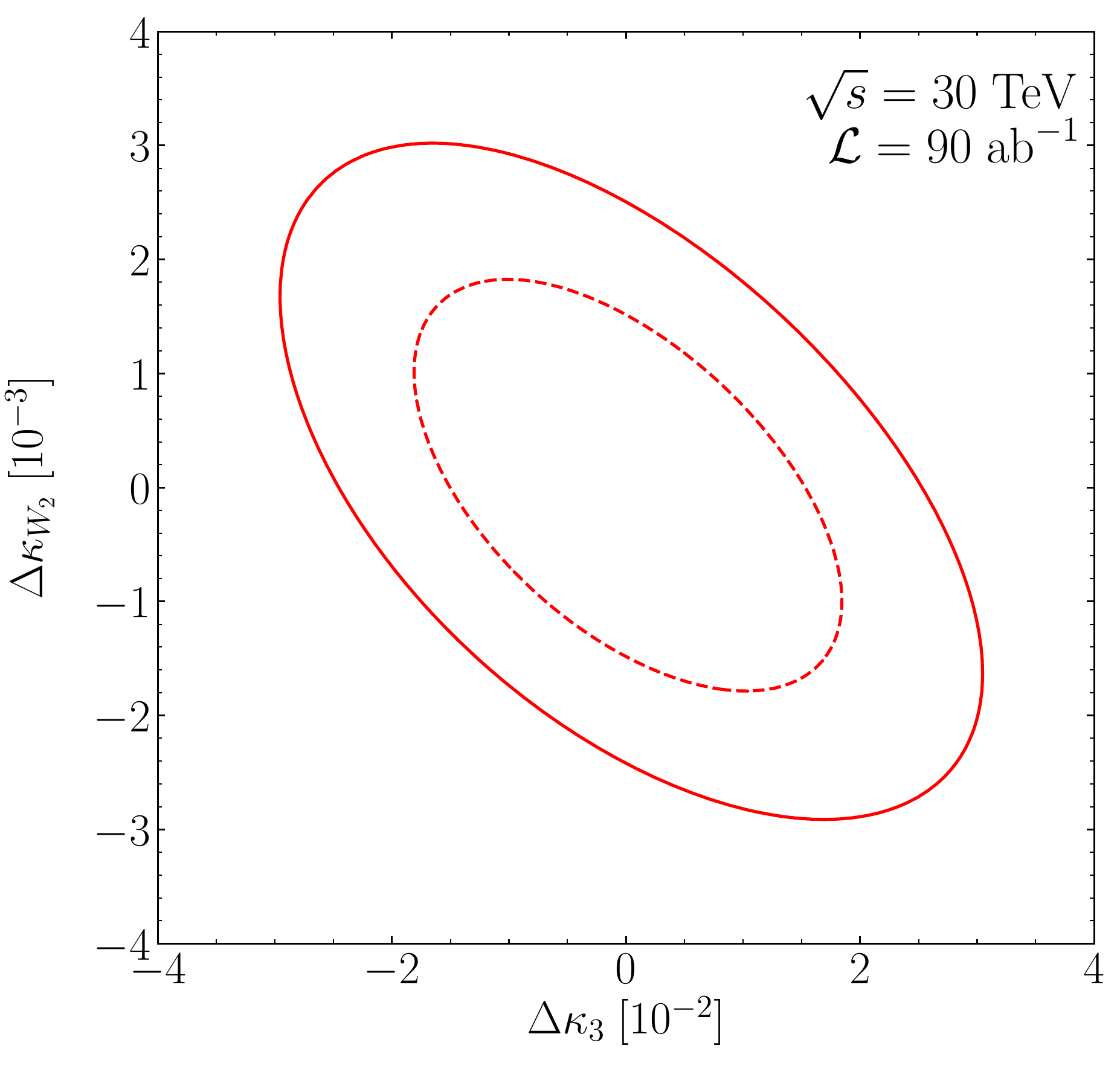}
\caption{}
\end{subfigure}
\caption{Correlated bounds with 95\% C.L.~(solid) and 68\% C.L.~(dashed) in the $\Delta\kappa_{W_2}$-$\Delta\kappa_3$ plane for $\sqrt s=3,6,10,30$ TeV, respectively. In (a), inner ellipses (solid) include the 95\% C.L.~results for 10 TeV and 30 TeV for comparison. }
\label{fig:hh_contours}
\end{figure}

We again perform a simultaneous fit to $\kappa_3$ and $\kappa_{W_2}$ using binned maximum likelihood fit. Given the different energy dependence in the subamplitudes controlled by $\kappa_3$ and $\kappa_{W_2}$, we decided to bin the $m_{HH}$ distribution into the following intervals\footnote{A similar procedure for double Higgs production in hadron colliders can be found in Ref.~\cite{Chen:2014xra}.}
\begin{equation}
m_{HH}^{} = [0, 350, 450, 550, 650, 750, 950, 1350, 5000]~{\rm GeV}.
\end{equation}
The binned cross section of $\mu^+\mu^-\rightarrow  HH + X \rightarrow b\bar{b}\ b\bar{b} + X$ after the selection cuts can be parametrized, in a similar fashion, as 
\begin{equation}
\sigma = \sigma_{\rm SM}\left[1 + r_1 \Delta\kappa_{W_2} + r_2\Delta\kappa_3 + r_3 \Delta\kappa_{W_2}\Delta\kappa_3 + r_4 \left(\Delta\kappa_{W_2}\right)^2 + r_5 \left(\Delta\kappa_3\right)^2\right],
\label{eq:rs}
\end{equation}
where the values are given in Table \ref{tab:rs10TeV} for $\sqrt s=10$ TeV for illustration. 
It is important to note again the increasing sensitivity on $\kappa_{W_2}$ at higher values of $m_{HH}^{}$. The resulting contours are shown in Fig.~\ref{fig:hh_contours}. In Table~\ref{tab:VVHHfit} we also provide the 95\% C.L. from the single parameter fit, by allowing $\kappa_3$ and $\kappa_{W_2}$ to vary only one at a time.

\section{Discussion and Conclusion}
\label{sec:Sum}

As we have shown in this work, a multi-TeV high energy muon collider will have a tremendous potential to constrain the electroweak Higgs couplings with  unprecedented accuracy. It will offer a unique probe into the nature of the Higgs boson as well as the scale of possible new physics beyond the SM. 
In Table \ref{tab:accurate}, we present a summary of the estimated sensitivities  at different collider energies and luminosities. In the last column of the table, we compare with the expected precision from other proposed colliders. It is clear that a multi-TeV muon collider could improve the measurements substantially. 

\begin{figure}[tb]
\centering
\includegraphics[width=0.8\textwidth]{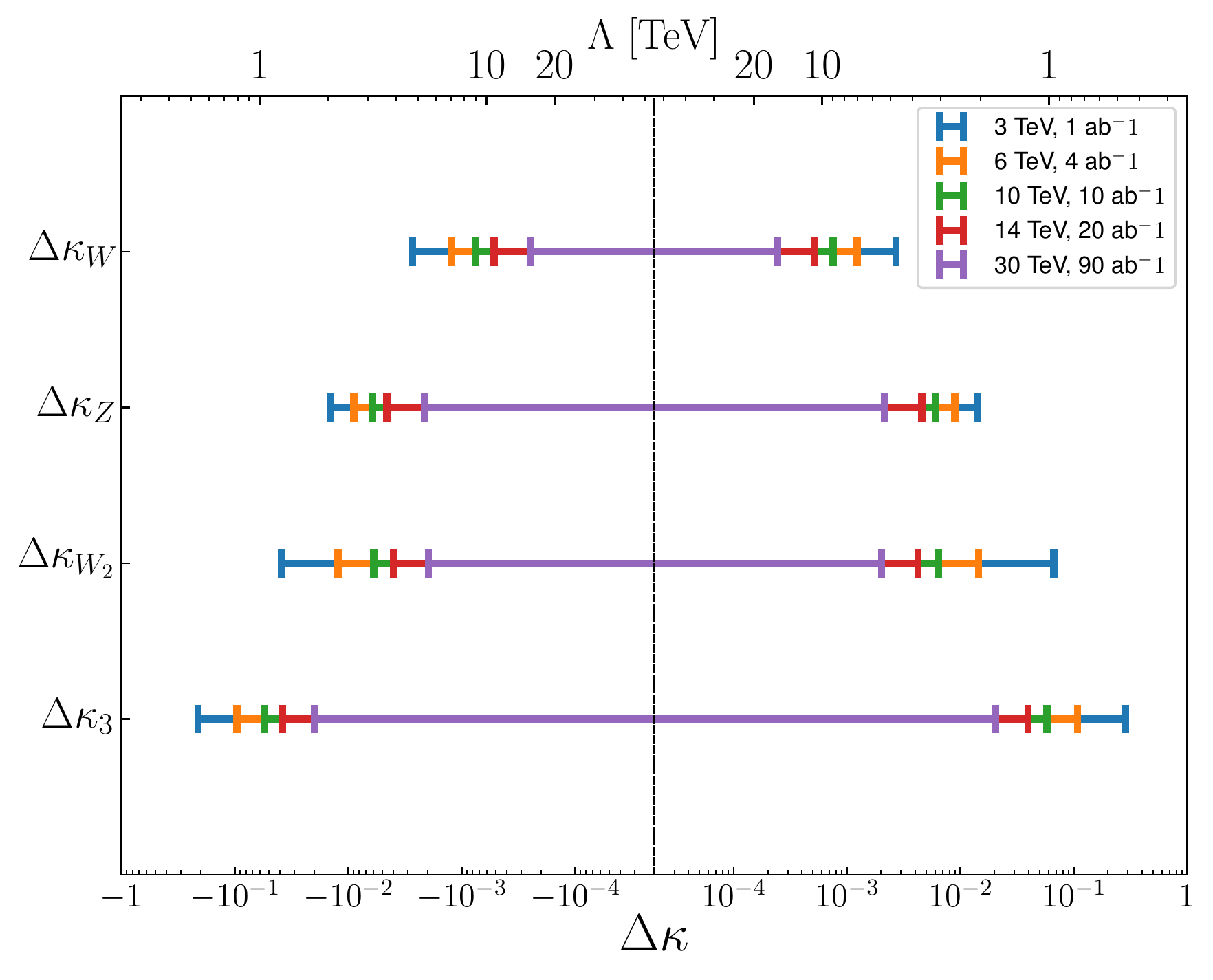}
\caption{Summary of the expected accuracies at $95\%$ C.L.~for the Higgs couplings at a variety of muon collider collider energies and luminosities. The upper horizontal axis marks the accessible scale $\Lambda$, assuming $c_{6,H}\sim {\cal O}(1)$. 
}
\label{fig:bars}
\end{figure}

It is possible to translate the bound in the $\kappa$-scheme into the constraint on $\Lambda$, the scale of new physics associated with the dim-6 operators in Eq.~(\ref{dim6text}),
\beq
\Lambda \sim \sqrt{{c^{}_{H,6}\over \Delta\kappa}} \ v.
\eeq
Assuming $c_{6,H}\sim {\cal O}(1)$, the scale is estimated to be $\Lambda \sim 1\ {\rm TeV}/\sqrt{16\Delta\kappa}$, as shown in Table \ref{tab:accurate}. 
A summary figure, which combines our results for the coupling measurements, is given in Fig.~\ref{fig:bars}, with the upper horizontal axis marking the estimated scale $\Lambda$ in TeV. With $\Lambda / \sqrt c_i \sim (10-16)$ TeV at a collider of ($10-30$) TeV, we would be probing new physics at very high scales or deeply into quantum effects.

\begin{table*}[th]
\centering
\begin{tabular}{|c|c|c|c|c|c||c|}
\hline
   $\sqrt s$~~ (lumi.)   & {3 TeV}  (1 ab$^{-1}$)  & {6} (4)& {10} (10)& {14} (20)& {30} (90) & Comparison\\
 \hline
   $WWH\ (\Delta\kappa_W)$ & 0.26\% &  0.12\%        &  0.073\%         & 0.050\%      &  0.023\% & 0.1\% 
 \cite{deBlas:2018mhx} \\
  $\Lambda/\sqrt c_i $ (TeV) & 4.7 &  7.0        &  9.0         & 11           &  16  &  (68\% C.L.)  \\
  \hline
   $ZZH\ (\Delta\kappa_Z)$ & 1.4\% &  0.89\%        &  0.61\%         & 0.46\%      &  0.21\% & 0.13\% 
      \cite{An_2019} \\
  $\Lambda/\sqrt c_i $ (TeV) & 2.1 &  2.6        &  3.2         & 3.6           &  5.3  &  (95\% C.L.) \\
  \hline
    $WWHH\ (\Delta\kappa_{W_2})$ & 5.3\% &  1.3\%        &  0.62\%         & 0.41\%      &  0.20\% & 5\%  \cite{Contino:2013gna} \\   
      $\Lambda/\sqrt c_i $ (TeV) & 1.1 &  2.1        &  3.1         & 3.8           &  5.5 &  (68\% C.L.)   \\
  \hline
       \hline
    $HHH\ (\Delta\kappa_3)$  & 25\% &  10\%        &  5.6\%         & 3.9\%      &  2.0\% & 5\% 
  \cite{Benedikt:2018csr,CEPC-SPPCStudyGroup:2015csa} \\
      $\Lambda/\sqrt c_i $ (TeV) & 0.49 &  0.77        &  1.0         & 1.2           &  1.7 &  (68\% C.L.)  \\
  \hline
\end{tabular}
\caption{Summary table of the expected accuracies at $95\%$ C.L.~for the Higgs couplings at a variety of muon collider collider energies and luminosities.
}
\label{tab:accurate}
\end{table*}
%

In our analyses, we only focused on the leading decay channel $H\to b\bar b$. A more comprehensive study could include the other decay channels as well, such as $H\to, WW, ZZ,\tau\tau$ and $\gamma\gamma$, to further improve the precision. On the other hand, due to the lack of knowledge of the specifics of the detector design, we have not made any attempts for experimental detector simulations. Further work may be needed to draw a more complete conclusion for the expected sensitivity reach. 

In summary, we estimated the expected precision at a multi-TeV muon collider for measuring the Higgs boson couplings with electroweak gauge bosons, $HVV$ and $HHVV$, as well as the trilinear Higgs self-coupling $HHH$. With the anticipated high CM energies and high luminosities,  a multi-TeV muon collider could provide us with unparalleled precision for Higgs physics and, consequently, offer some of the most stringent experimental tests of the SM Higgs sector. 
As we have shown in this study, the outgoing remnant particles have a strong tendency to stay in the very forward region.
The enhanced collinear behavior of the final state particles results in the dominant configuration of  ``inclusive'' processes, a notion usually reserved for hadron colliders, unless there is a device to detect the very forward muons of a few degrees from the beam. 
These features add new subtlety to Higgs coupling measurements, since it is now difficult to isolate $WW$ fusion from $ZZ$ fusion events in the Higgs production. We addressed the subtlety by performing binned maximum likelihood analyses to simultaneously fit two parameters involved in the inclusive processes. The approach and  methodology adopted in this study could be applicable to new physics searches at a high energy muon collider.

\begin{acknowledgments}
The work of TH was supported in part by the U.S.~Department of Energy under grant No.~DE-FG02- 95ER40896 and in part by the PITT PACC. The work of DL was supported in part by the U.S. Department of Energy under grant DE-SC-0009999.  IL is supported in part by the U.S. Department of Energy under contracts No. DE- AC02-06CH11357 at Argonne and No. DE-SC0010143 at Northwestern.  XW was supported by the National Science Foundation under Grant No.~PHY-1915147.
 \end{acknowledgments}

\bibliographystyle{utphys} 
\bibliography{refs}

\end{document}